\documentclass[fleqn,usenatbib]{mnras}

\usepackage{newtxtext,newtxmath}

\usepackage[T1]{fontenc}

\DeclareRobustCommand{\VAN}[3]{#2}
\let\VANthebibliography\thebibliography
\def\thebibliography{\DeclareRobustCommand{\VAN}[3]{##3}\VANthebibliography}

\defcitealias{Vinkovic21}{V\v{C}21}

\usepackage{graphicx}	
\usepackage{amsmath}	

\title[The role of non-radial radiation pressure]{Inner dusty regions of protoplanetary discs - III. The role of non-radial radiation pressure in dust dynamics}

\author[D. Vinkovi\'{c} and M. \v{C}emelji\'{c}]{
Dejan Vinkovi\'{c},$^{1,2}$\thanks{E-mail: dejan@iszd.hr (DV), miki@camk.edu.pl (M\v{C})}
and Miljenko \v{C}emelji\'{c},$^{3,4,5,6}$
\\
$^1$Science and Society Synergy Institute, Bana Josipa Jela\v{c}i\'{c}a 22,HR-40000, \v{C}akovec, Croatia\\
$^2$Oraclum Intelligence Systems Ltd., 23 Arnold Close, Hauxton, Cambridge CB22 5FN, UK\\
$^3$SGMK Nicolaus Copernicus Superior School, College of Astronomy and Natural Sciences, Nowogrodzka 47A, 00-697, Warsaw, Poland\\
$^4$Research Centre for Computational Physics and Data Processing, Institute of Physics, Silesian University in Opava, Bezru\v{c}ovo n\'am.~13, CZ-746\,01, Opava,\\ Czech Republic\\
$^5$Nicolaus Copernicus Astronomical Center, Polish Academy of Sciences, Bartycka 18, 00-716 Warsaw, Poland\\
$^6$Academia Sinica, Institute of Astronomy and Astrophysics, P.O. Box 23-141, Taipei 106, Taiwan
}

\date{Accepted .... Received ...; in original form ...}

\pubyear{2023}

\begin{document}
\label{firstpage}
\pagerange{\pageref{firstpage}--\pageref{lastpage}}
\maketitle

\begin{abstract}
We explore dynamical behaviour of dust particles that populate the surface of inner optically thick protoplanetary discs. This is a disc region with the hottest dust and of a great importance for planet formation and dust evolution, but we still struggle to understand all the forces that shape this environment. In our approach we combine results from two separate numerical studies - one is the wind velocity and density distributions obtained from magnetohydrodynamical simulations of accretion discs, and the other is a high-resolution multigrain dust radiation transfer. In our previous paper in the series, we described the methodology for utilising these results as an environmental input for the integration of dust trajectories driven by gravity, gas drag, and radiation pressure. Now we have two improvements - we incorporate time changes in the wind density and velocity, and we implement the non-radial radiation pressure force. We applied our analysis on Herbig Ae and T Tau stars. We confirm that the radiation pressure force can lead to dust outflow, especially in the case of more luminous stars. Additionally, it opposes dust accretion at the inner disc edge and reduces dust settling. These effects are enhanced by the disc wind, especially in the zone where the stellar and the disc magnetic fields meet. Our results suggest that dust grains can stay in the hottest disc region for an extended period and then can end up ejected into the outer disc regions.
\end{abstract}




\section{Introduction}

The inner edge of protoplanetary discs, situated in close proximity to the central star, represents a critical zone where complex processes prepare the stage for formation of planetary systems. This is the hottest part of the disc where the gas accretes onto the central star, but also flows outward as a low density wind. The gas is accompanied by dust, up to the point where grains sublimate. The whole dust and gas dynamics in this region is exposed to various forces, turbulence, and instabilities, all intertwined in a complicated way \citep{Lesur}.

This zone also has a small angular size, on the top of being in the immediate vicinity of a bright central star. Hence, it is difficult to explore inner disc regions not only theoretically, but also observationally \citep{Review2010}. 
The approach taken so far is to identify the key forces that shape the disc and then disentangle the processes that drive the gas and dust dynamics, and infer how they impact the observations \citep[e.g.][]{Vinkovic14,Flock17}. Dust transport and evolution is of a special interest in this context as an initial ingredient of the planetary systems formation. High densities and temperatures alter the grains' physical and chemical properties, while intricate details of dust dynamics can transport this processed dust into outer disc regions \citep{Testi2014}.

In recent years interaction between gaseous disc winds and dusty discs have been explored as a method for expanding possibilities for dust dynamics. Disc wind might lift dust higher above the disc than the typical dusty disc scale heights and transport dust from inner to outer regions \citep{Pascucci}. Ubiquitous variability in young pre-main-sequence stars is often interpreted as episodes of dust clouds dimming the star \citep[e.g.][]{Cody2014,Rice2015,Capistrant2022}, which might also be driven by disc winds. 
Interestingly, much less attention has been given to the role of radiation pressure, which is one of the fundamental forces of light-dust interaction. It has been shown that this force has a significant impact on the dust populating the disc surface where grains are exposed to the direct stellar radiation \citep{Takeuchi,Vinkovic09}. 

This motivated us to explore in more detail what that means for the dynamics of dust in the inner disc regions. We also wanted to be more realistic in the description of conditions that dictate the environment in which the inner disc dust exists and moves. Hence, in our previous paper \citep[][hereafter: \citetalias{Vinkovic21}]{Vinkovic21} we presented a methodology that combines results from a high-resolution multigrain radiative transfer in dusty discs with disc winds obtained from magnetohydrodynamical (MHD) simulations. This enables us to use a more realistic mapping of radiation pressure force and use a more insightful behaviour of gas density and velocity in the wind. 

In \citetalias{Vinkovic21} we applied that methodology in a limited way. We used only a snapshot from the MHD simulation, thus avoiding time-dependent properties of environmental variables in dust dynamics equations. We also looked only at the case of T Tau stars, where the stellar luminosity, dictating the strength of radiation pressure force, is low. Nonetheless, results were very interesting and insightful as we showed how dust can undergo various dynamical scenarios. In some cases, dust can be ejected, while in other it can be trapped, opposing accretion toward the star. We demonstrated the importance of radiation pressure force as an unavoidable component in the intricate interaction between the stellar radiation and the material constituting the protoplanetary disc. 

In this paper we go further. Here we include time changes in the wind density and velocity structure and adjust the equations to accommodate this addition. We also look at Herbig Ae stars, whose higher luminosity boosts the importance of radiation pressure. Another improvement is addition of the non-radial radiation pressure force caused by the radiation coming from the hot inner disc dust. Since big grains can survive closest to the star, they are populating the surface of optically thick inner disc edge. \cite{Vinkovic09} showed that, because of their grey opacity in visual and near-infrared wavelengths, these grains experience push not only from the stellar radiation, but also from the hot dust itself. The net result is a force tangential to the disc surface that can push the grains out of the disc surface into the regions of lower gas density above the disc where the strong stellar radiation pressure force takes over and ejects grains outward. In this paper we show how this affects the dust dynamics in Herbig Ae and T Tau stars.

We start in Section \ref{SecHerbigAe} with a Herbig Ae star and show how our methodology is adapted to this case, how we solve the dust dynamics equations, and what we see in the results. In Section \ref{SecTTAu} we do the same for a T TAu star. We discuss results in Section \ref{SecDiscuss} and summarise them in Section \ref{SecConclude}.

\section{Dusty outflow in Herbig Ae stars}
\label{SecHerbigAe}

The work presented here builds upon our previous exploration of the structure of optically thick dusty discs around Herbig Ae stars. The key preliminary results on the dust dynamics effects of non-radial radiation pressure force were presented by \cite{Vinkovic09}. We will now go into more detail using the radiation pressure forces calculated in that study. The intricate details of the optically thick dusty disc used in our model were presented by \cite{Vinkovic12}. The details were derived using a high-resolution radiative transfer calculation capable of resolving the large temperature gradients and disc-surface curvatures
caused by dust sublimation. Now we can use only a subset of these results that are relevant for the dust dynamics on the surface of dusty disc at the inner disc edge. The numerical approach is based on the dust dynamics equations presented by \citetalias{Vinkovic21}. 

\subsection{Description of the model}

The focus of this work is analysis of the impact of radiation pressure on the movement of dust particles in optically thick dusty discs. We can expect such discs to be highly complicated environments exposed to a variety of forces. Therefore, we need to simplify the disc model to the level that we can efficiently simulate, but still achieve the main goal: exploring caveats of non-radial radiation pressure effects on the dynamics of dust that populates the surface of inner disc edge. 

In Herbig Ae stars we are helped by the high luminosity of those stars. In our model the star has $L_*=35L_\odot$ and the temperature of \hbox{$T_*=$10~000~K}. We expect the accretion luminosity to be much smaller and originating mostly close to the star, where the inner edge of our dusty disc is at the distance of $R_\text{in}=44.28R_*=0.412$~AU from the star, while the stellar radius is $R_*=2R_\odot$ and the stellar mass is $M_*=2M_\odot$. That distance is large enough that we do not need to consider disc winds. Such disc winds are more pronounced closer to the star as we will see in the next section where we model discs around T Tau stars using MHD simulations. 

The simplified vertical disc structure is described with the Shakura–Sunyaev model \citep{SS73}, where we also assume that the gas and dust are well mixed. This gives the following number density of dust particles in the cylindrical coordinate system ($\varrho$,$z$):
\begin{equation}
    \label{EqDisc}
    n_\alpha(\varrho,z)=N_\alpha D(\varrho,z)=N_\alpha (\varrho/R_\text{in})^{-p}\exp\left( {-h_0 \frac{(z/R_\text{in})^2}{(\varrho/R_\text{in})^{2m}}}\right)
\end{equation}
where $\alpha$ is the dust type, $N_\alpha$ is the number density of dust grains at ($R_\text{in}$,0), and the parameters are $p$=2, $m$=1.25 and $h_0$=1800. 

In our previous work \citep{Vinkovic12} we described the radiative  transfer solution for this disc structure with a mix of two dust types: 0.1~$\mu$m radius representing small grains, and 2~$\mu$m radius representing big grains. Small grains are more numerous but evaporate more easily than big grains. Hence, only the big grains populate the hottest inner edge of the dusty disc. Small grains exist within the optically thick disc interior behind a few visual optical depths, where they are shielded by big grains from direct stellar radiation. How this impacts disc shape and dust distribution is described by \cite{Vinkovic12}, including the values of $N_\alpha$. The optical depth along a spatial step $dl$ consistent with this density model is
\begin{equation}
    \label{EqTau}
    d\tau(\varrho,z)=790 \left( \frac{R_*}{R_\text{in}}\right) D(\varrho,z) dl(\varrho,z)\, .
\end{equation}

The key aspect of multigrain high-resolution radiation transfer calculation is the realisation that big grains populate the hot inner edge, but at the same time their peak temperature is within the optically thick disc \citep[see Figs. 4 and 5 in][]{Vinkovic12}. In other words, big grains can move closer to the star than the inner edge as long as this extended dusty zone is optically thin to the stellar radiation. Therefore, the whole inner disc can be surrounded by an optically thin cloud of big grains. For a more detailed analysis of this effect we recommend the work by \cite{Kama09} \citep[see also][]{Flock16}, while a general theoretical proof that the effect is real and not a numerical artefact is given by \cite{Vinkovic06}.

The question that we explore here is how the radiation pressure force affects grains in the inner disc surface and in the optically thin halo. Without any further analysis we know that grains that are not constrained by the gas drag are influenced primarily by gravity and radiation pressure. The high luminosity of Herbig Ae stars results in ejection of such dust when their radius is $\lesssim$10~$\mu$m. An open question is what happens to dust trajectories in the disc surface where the gas drag force becomes non-negligible. 

\cite{Takeuchi} explored this problem using only the force of stellar radiation pressure. They concluded that surface dust is pushed slightly outward, but grains remain constrained to the disc surface. Essentially, the surface dust is pushed deeper into the disc, where the grains are shielded from the strong stellar radiation and/or the gas drag dominates. 

However, there is a twist to this story. \cite{Vinkovic09} showed that in the case of big grains (sizes $\gtrsim$1~$\mu$m) we cannot ignore the radiation pressure force created by the dust thermal radiation. The temperatures at the inner dusty disc edge are in the sublimation range of silicates (or even higher in case of more resilient dust), which is $\sim$1500~K. The peak of thermal radiation from this hot dust is in the near-infrared wavelength range. Since the big grains are of sizes similar or bigger than these wavelengths, the radiation pressure force becomes significant when compared to the gravity force. 

We can write the radiation pressure force vector $\bmath{\mathfrak{F}}$ as a sum of stellar component $\bmath{\mathfrak{F^*}}$ (includes also accretion, if it is not negligible compared to the stellar luminosity) and diffuse (thermal) component $\bmath{\mathfrak{F^\text{diff}}}$. For simplicity we scale these values with the pure stellar radiation pressure and get a scaled radiation pressure force vector
\begin{equation}
    \label{EqRadPressVec}
    \bmath{\mathfrak{f}}(\varrho,z)=\frac{\bmath{\mathfrak{F^*}}+\bmath{\mathfrak{F^\text{diff}}}}{\int\sigma_\lambda\text{F}^*_\lambda d\lambda/c}=\mathfrak{f_r}\bmath{\hat{r}}+\mathfrak{f_\theta}\bmath{\hat{\theta}}\,\,,
\end{equation}
where $\sigma_\lambda$ is the extinction cross section of dust grains, $\text{F}_\lambda^*$ is the stellar flux, $c$ is the speed of light, and ($\bmath{\hat{r}}$,$\bmath{\hat{\theta}}$) are the spherical coordinate unit vectors. We see that far away from the disc ($\mathfrak{f_r}$,$\mathfrak{f_\theta}$)=(1,0), while deep within the disc ($\mathfrak{f_r}$,$\mathfrak{f_\theta}$)=(0,0). 

Unfortunately, calculating $\bmath{\mathfrak{f}}$ is a highly non-trivial task. The diffuse radiation originates from the entire inner region of dusty disc, which makes the pressure vector pointing in a non-radial direction. Hence, we will use results from the high-resolution radiative transfer calculation by \cite{Vinkovic09}. This calculation was performed for the same disc structure as in equation \ref{EqDisc}. The force was acting on olivine grains of 2~$\mu$m in radius, with the optical properties taken from \cite{Dorschner95}.

\begin{figure}
 \includegraphics[width=\columnwidth]{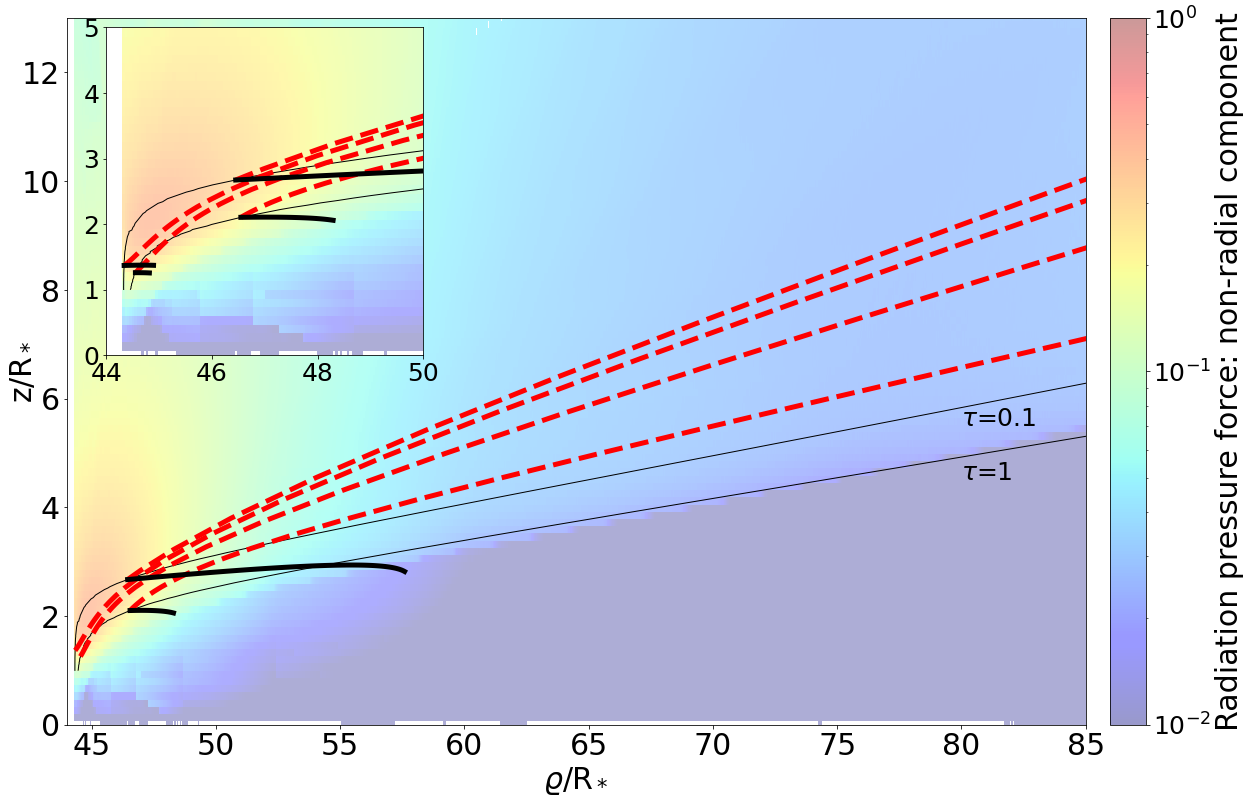}
 \caption{The figure shows examples of dust dynamics with and without the influence of non-radial radiation pressure force in a Herbig Ae disc. The surface of an optical thick dusty disc is marked by two solid thin lines - one tracing the radial visual optical depth of $\tau=1$ and another tracing $\tau=0.1$. The colour map shows the strength of non-radial radiation pressure relative to the value that the radial radiation pressure would have if there were no dust ($\mathfrak{f_\theta}$ in equation \ref{EqRadPressVec}). The thick dashed lines are examples of trajectories of four particles released from the inner disc edge. Their full 3D-path is shown in Fig. \ref{fig:HAeFig2}, while here we show the change of cylindrical coordinates $\varrho$ and $z$ scaled with the stellar radius $R_*$. The thick solid black lines are trajectories of the same dust particles when the non-radial radiation pressure force is not taken into consideration, i.e. when only the radial force is active. The inset chart shows the same situation, but zoomed on the inner disc edge.  }
 \label{fig:HAeFig1}
\end{figure}

Interesting results stem from the non-radial component $\mathfrak{f_\theta}$, shown as the colour map in Fig.~\ref{fig:HAeFig1}. In order to explore the impact of $\mathfrak{f_\theta}$ on dust trajectories, we need to incorporate the "strength" of the radiation pressure force into the dust dynamics equations. This strength is measured by the ratio of the radiation pressure and the local gravity force. It can be expressed as \citep{Vinkovic09}
\begin{equation}\label{EqBeta}
\bmath{\beta}(\varrho,z)=0.19 \left(\frac{L_*}{L_\odot}\right)\left(\frac{M_\odot}{M_*}\right)
\left(\frac{3000 \, \text{kg}\, \text{m}^{-3}}{\rho_{\mathrm grain}}\right)\left(\frac{\mu \text{m}}{a}\right) \langle Q\rangle_{*} \bmath{\mathfrak{f}}(\varrho,z),
\end{equation}
where $\rho_{\mathrm{grain}}$ is the grain's bulk density, $a$ is the grain's radius, and $\langle Q\rangle_{*}$ is the extinction coefficient $Q_\lambda$ averaged over the stellar spectrum shape $f_\lambda^*=\text{F}_\lambda^*/\int \text{F}_\lambda^*d\lambda$
\begin{equation}
\label{Qstar}
\langle Q\rangle_{*} = \int \frac{\sigma_\lambda}{a^2\pi}  f_\lambda^*  d\lambda=\int Q_\lambda  f_\lambda^*  d\lambda\,\,.
\end{equation}

\subsection{Calculating dust trajectories}
\label{Sec:HAe_trajectory}

Equations of motion for a single dust particle at position $\bmath{r}=(\varrho,\varphi,z)$ in the cylindrical coordinate system are derived by \citetalias{Vinkovic21} using scaled variables $r\equiv r/R_*$, $t\equiv t/t_0$ and $\bmath{V}\equiv \bmath{V}\,t_0/R_*$, where $t_0=(R_*^3/GM_*)^{0.5}$. The gas velocity is fixed to the Keplerian velocity $\bmath{V}_{gas}=(\text{v}_{,{\mathrm gas},\varrho},\text{v}_{{\mathrm gas},\varphi},\text{v}_{{\mathrm gas},z})=(0,\varrho^{-0.5},0)$ in scaled units. The equations that we must solve are
\begin{equation}\label{eq:ForceRho}
\ddot{\varrho}=\varrho \dot{\varphi}^2 - \frac{\varrho}{(\varrho^2+z^2)^{3/2}} - \mu \dot{\varrho}+\frac{\beta_\varrho}{\varrho^2+z^2}
\end{equation}
\begin{equation}\label{eq:ForcePhi}
\ddot{\varphi}=-2\frac{\dot{\varrho}}{\varrho}\dot{\varphi}-\mu \left(\dot{\varphi}-\frac{\text{v}_{{\mathrm gas},\varphi}}{\varrho}\right)
\end{equation}
\begin{equation}\label{eq:ForceZ}
\ddot{z}= - \frac{z}{(\varrho^2+z^2)^{3/2}} - \mu \dot{z} +\frac{\beta_z}{\varrho^2+z^2} ,
\end{equation}
where the radiation pressure vector from equation \ref{EqBeta} is $\bmath{\beta}=(\beta_\varrho,0,\beta_z)$. In the case of Herbig Ae stars with the disc structure from equation \ref{EqDisc}, we combine  Section 5.3 from \citetalias{Vinkovic21} with the disc density from \cite{Vinkovic12} and derive
\begin{equation}\label{eq:mu}
\mu(\varrho,z)=\frac{6.07\times 10^{-4}}{a}
\left(\frac{R_\odot}{R_*}\right)\, t_0\,
T^{1/2}(\varrho,z)\,  D(\varrho,z),
\end{equation}
where $a$ is in the units of $\mu m$ and $T$ is the gas temperature. 

The model variables dependent on spatial coordinates are precalculated over a high-resolution grid to simplify and speed up the simulation. The sequence of computational steps is as follows:
\begin{enumerate}
\item Set the grain size $a$=2$\mu$m and import its optical properties $Q_\lambda$.
\item Calculate $\langle Q\rangle_{*}$ from equation \ref{Qstar}.
\item Calculate the disc density profile $D(\varrho,z)$ from equation \ref{EqDisc} over the grid. 
\item Calculate the Keplerian gas velocity $\text{v}_{{\mathrm gas},\varphi}$ over the grid. 
\item Calculate an array of optical depths using equation \ref{EqTau} over the grid. 
\item Import the scaled radiation pressure force vectors $\mathfrak{f_r}$ and $\mathfrak{f_\theta}$, and transform them into $\bmath{\beta}(\varrho,z)$ using equation \ref{EqBeta}.
\item Calculate temperature over the grid using approach from Section 5.2 by \citetalias{Vinkovic21}.
\item Calculate an array of $\mu(\varrho,z)$ using equation \ref{eq:mu} over the grid.
\item Define the initial position of a dust grain and set the initial velocity using the steady state equations described in Section 5.5 by \citetalias{Vinkovic21}.
\item Integrate equations \ref{eq:ForceRho}-\ref{eq:ForceZ} to follow the dust grain trajectory. 
\end{enumerate}

For now, we are ignoring possible dust evaporation when smaller grains exit the optically thick disc interior or pass through gas layers of higher temperatures. While big grains can survive this process in this disc region, grains can also undergo various other physical and chemical alternations (partial sublimation, collisional fragmentation, fractal growth, crystallisation, charging, etc.). 

\begin{figure}
 \includegraphics[width=\columnwidth]{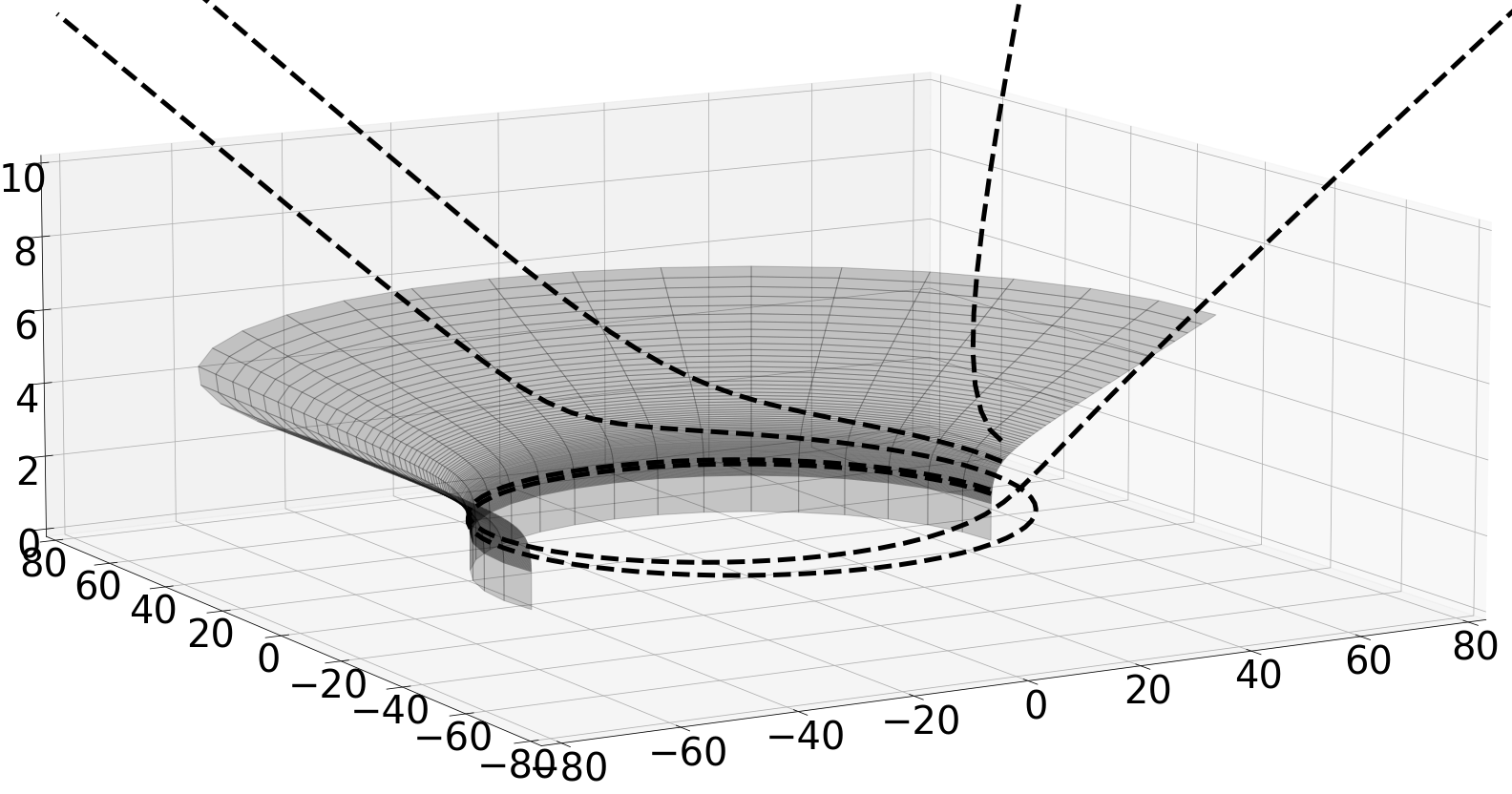}
 \caption{The same dynamical configuration as Fig. \ref{fig:HAeFig1}, but shown in 3D Cartesian coordinates scaled with the stellar radius. The 3D surface shows the location of radial visual optical depth of $\tau=1$. Four particles released from different heights at the inner disc edge take different time spans circling the star. Their paths are shown as dashed lines. Particles at lower heights take longer time to climb up and then get ejected. This is further illustrated in Fig. \ref{fig:HAeFig3}.}
 \label{fig:HAeFig2}
\end{figure}

\begin{figure}
 \includegraphics[width=\columnwidth]{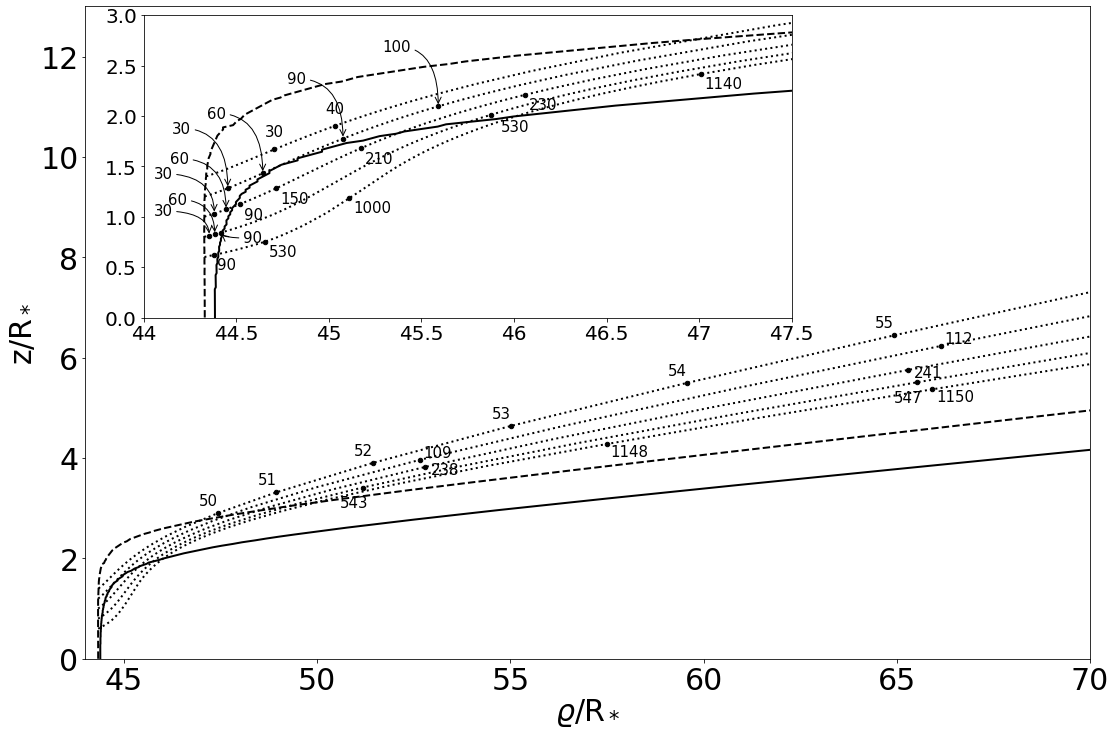}
 \caption{Trajectories of dust particles released from different locations at the inner dusty disc edge around a Herbig Ae star. The thick dashed and solid lines trace the radial optical depths of $\tau=0.1$ and $\tau=1$, respectively. The dotted lines show paths of various dust particles. The plot follows only their $\varrho$ and $z$ cylindrical coordinates scaled with the stellar radius $R_*$, while the whole 3D path is illustrated in Fig. \ref{fig:HAeFig2}. Dots with numbers show the time in days that it takes for a particle to reach that location after the initial release. For comparison, the orbital period at the inner dusty disc radius is 67 days. The inset chart shows the same situation, but zoomed on the inner disc edge. The plot illustrates the ability of non-radial radiation pressure force to erode the inner dusty disc edge over longer periods of time and deposit this dust into the outer parts of the disc \citep{Vinkovic09}.}
 \label{fig:HAeFig3}
\end{figure}

\begin{figure*}
 \includegraphics[width=2\columnwidth]{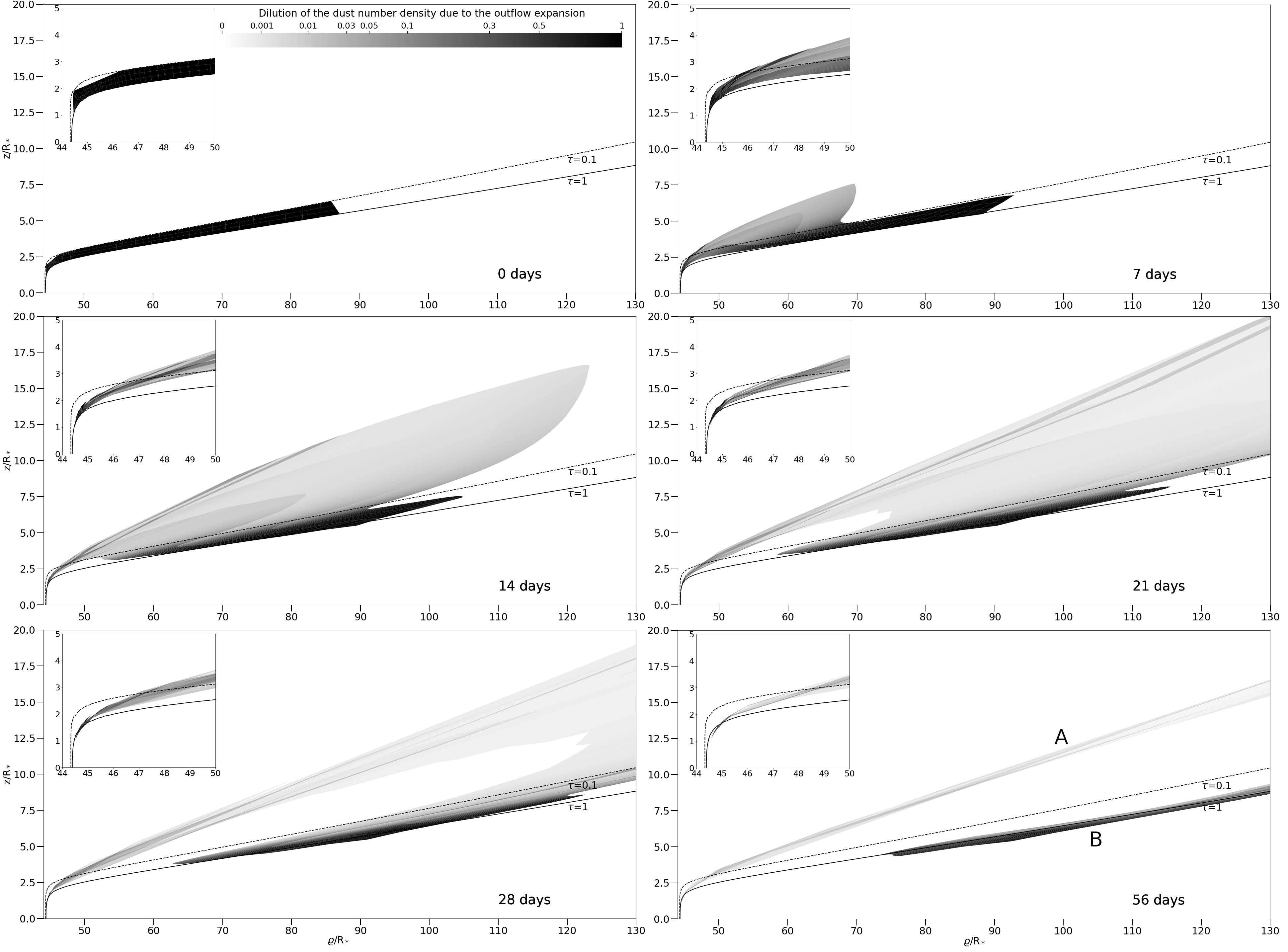}
 \caption{A time sequence of dusty outflow driven by the non-radial radiation pressure force. The dashed and solid lines trace the radial visual optical depth of $\tau=0.1$ and $\tau=1$, respectively, in an optically thick dusty disc around a Herbig Ae star. The dust particles are initially positioned within the black zone shown in the top left plot, with the initial velocity that follows the rotation of the gas disc. 
 Particles are grouped into groups of four that form Lagrangian cells in cylindrical coordinates ($\varrho$,$z$) scaled with the stellar radius $R_*$. Hence, the dynamical evolution of individual particles is mapped into time evolution of the Lagrangian cells. Plots are snapshots of this outflow evolution at times indicated as days in the plots. As the cells grow in size, the dust number density within the cells drops. The greyscale colour map shows this dilution as the volume of cells scaled with the initial cell volume on the day zero. The inset chart shows the same situation, but zoomed on the inner disc edge. In the last plot (bottom right) two final dust outflow configurations are marked: the letter $A$ shows the persistent erosion of the inner disc edge driven by the non-radial radiation pressure  \citep{Vinkovic09}, while $B$ marks the dust pushed by the radial radiation pressure force deeper into the disc surface \citep{Takeuchi}. 
Animation showing the outflow is available in the supplementary materials as \texttt{Figure\_4.mp4} (see Appendix \ref{AppendixAnima}).}
 \label{fig:HAeFig4}
\end{figure*}

\subsection{Results}

We position dust grains into the surface layer of inner disc edge where we see that the non-radial component $\mathfrak{f_\theta}$ of the radiation pressure force might have the largest impact on the grain trajectories. Fig. \ref{fig:HAeFig1} shows $\mathfrak{f_\theta}$ as colour map and the disc surface as lines of radial visual optical depth of $\tau=0.1$ and $\tau=1$. We select two illustrative starting points for dust at $\tau=0.1$ and two at $\tau=1$. Their dynamical evolution is shown in Fig. \ref{fig:HAeFig1} as paths tracing the change of their cylindrical coordinates $(\varrho,z)$. 

Ignoring $\mathfrak{f_\theta}$ leads to trajectories that push these grains deeper into the disc (thick solid black lines), where the optical depth is large enough to make the radiation pressure force irrelevant. This is the result obtained by \cite{Takeuchi} and it has its own interesting consequences for the inner disc edge. It leads to removal of dust from the inner edge surface, which means that the dusty disc is truncated at some height dictated by the balance between gas drag and radiation pressure on dust grains \citep{Vinkovic14}.

This dust removal becomes more interesting when the non-radial correction $\mathfrak{f_\theta}$ is included into the grain dynamics. The thick dashed lines in Fig.~\ref{fig:HAeFig1} show the tracks of selected grains. They are lifted out of the disc surface and transported outward, where they exit our computational domain. They are deposited in outer regions of the disc, which was discussed by \cite{Vinkovic09}. Interestingly enough, even grains at lower disc heights are extracted from the inner rim surface and transported outward. 

To understand better the dynamics of grain transportation, we need to consider the entire trajectory, including the azimuth coordinate. Fig. \ref{fig:HAeFig2} shows how these same four grains actually travel in 3D space. Grains at higher disc heights, where the gas drag is small, are easily pushed outward by the radiation pressure force. Grains embedded into the inner disc edge surface at lower disc heights, where the gas drag is stronger, take some time to climb up. When they reach the heights of lower gas drag conditions, they start to drift outward. 

We see from this that the timescales of dust outflow differ significantly between grains of different disc heights. Fig. \ref{fig:HAeFig3} shows this for several grains of different initial height positions within the inner edge. The time in days required for grains to reach various points on their trajectory is marked as numbers. Grains at lower heights, where the gas drag is strong, take many orbital periods before they reach the upper disc heights and get ejected (the Keplerian orbital period is 67 days at the inner disc rim in our case). Even though in reality dust will be affected by other forces over longer periods of time, we see that the non-radial radiation pressure force opposes dust settling within the surface of inner disc edge. 

Our next goal was to map the dust outflow from the entire surface of the inner edge. Since this would require a huge number of individual particles, we preferred the approximation where we group four neighbouring particles into a Lagrangian cell. The evolution of these four particles is then treated as the Lagrangian flow of particles within the cell. Fig.~\ref{fig:HAeFig4} shows that outflow evolution. The dust from inner edge disc surface, where $\mathfrak{f_\theta}$ is the strongest (see Fig.~\ref{fig:HAeFig1}), is lifted up and pushed toward the outer disc regions  \citep{Vinkovic09}. A tenuous narrow outflow of dust originating from the lower disc heights continues after the rest of the surface dust is already blown away (see the bottom right panel in Fig.~\ref{fig:HAeFig4}). The dust outside the strong impact of $\mathfrak{f_\theta}$ is pushed only by the radial radiation pressure force and, therefore, gets deposited deeper into the nearby surface \citep{Takeuchi}.  

\section{T~Tau stars}
\label{SecTTAu}

Since T~Tau stars have a much smaller luminosity than Herbig Ae stars, the dust can exist closer to the star, where gas disc wind can strongly interact with the dusty disc. We model the wind using MHD simulation and combine it with the dust dynamics as prescribed in our previous work \citepalias{Vinkovic21}, but this time with temporal changes in the disc wind. In this section we first describe the MHD model and rationale behind its implementation. Then we show how the dust dynamics theory is now improved, followed by the results of dust trajectory modelling. 

\begin{figure*}
 \includegraphics[width=2\columnwidth]{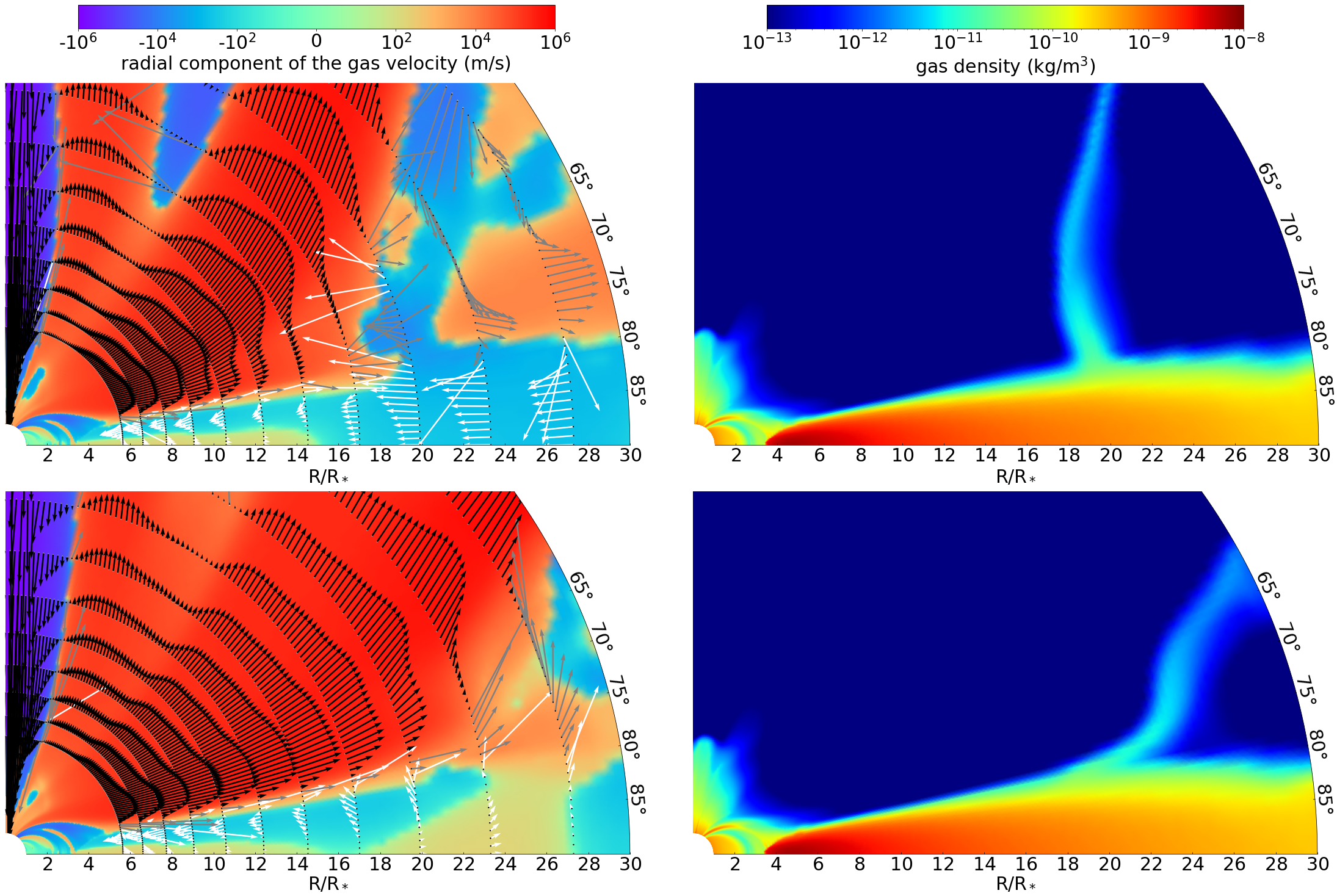}
 \caption{Two time instances (top and bottom panels) of the MHD model used in our study of T~Tau star. Left panels: The radial component of gas velocities (shown as the colour map) and the vectors of gas velocities. The black arrows depict velocities $>2\times 10^4$m~s$^{-1}$, the grey arrows are enlarged by a factor of 50 and they represent velocities between $2\times 10^4$m~s$^{-1}$ and $2\times 10^3$m~s$^{-1}$, and the white arrows are enlarged by a factor of 500 and they represent velocities $<2\times 10^3$m~s$^{-1}$. For comparison, the Keplerian velocity is 2.18$\times$10$^5$($R/R_*)^{-0.5}$m~s$^{-1}$. Right panels: The gas density of disc wind MHD model. Animations showing gas velocity and density changes are available in the supplementary materials as \texttt{Figure\_5\_velocity.mp4} and \texttt{Figure\_5\_density.mp4} (see Appendix \ref{AppendixAnima}). }
 \label{fig:MHD}
\end{figure*}

\subsection{Vertical perturbation in the disc corona}
\label{SubSec_disccorona}

\cite{Cem19} and \cite{CemBrun23} presented results of numerical simulations incorporating a thin protostellar disc in a parameter study which included 64 simulations with a slowly rotating star\footnote{The values for the stellar rotation rate were $\Omega_\star\le 0.2\Omega_{\mathrm br}$, where $\Omega_{\mathrm br}$ is the breakup velocity at the stellar equator.} and a range of resistivity coefficients $\alpha_{\mathrm m}\in (0.1,0.4, 0.7, 1.0)$ and stellar magnetic field strengths B$_\star\in (250,500,750,1000)$~G. In all 64 simulations the viscosity was defined through the anomalous alpha coefficient $\alpha_{\mathrm v}=1$, to avoid backflow in the midplane of the disc. Such a backflow appears in thin disc analytical or semi-analytical solutions and is confirmed by simulations in both purely hydro-dynamical and magnetic cases and also with alpha-prescription and magneto-rotational instability (MRI) \cite[and references therein]{Mishra}. Such a flow near the disc mid-plane, going against the overall accretion flow in the disc, was usually considered a numerical artefact or a consequence of inclusion of only a part of the viscous tensor. However, in simulations amending the \cite{Cem19} parameter study with different values of viscous coefficient $\alpha_{\mathrm v}$, \cite{Mishra} showed that the mid-plane backflow is of physical origin. For the (turbulent) magnetic Prandtl numbers P$_{\mathrm m}=2\alpha_{\mathrm v}/(3\alpha_{\mathrm m})<0.6$ they obtained intermittent or stable backflow in the disc mid-plane for slowly rotating stars with $\Omega_\star=0.1\Omega_{\mathrm br}$.

Incorporating dust dynamics into already a large variety of gas dynamics in MHD wind solutions requires a comprehensive investigation that goes beyond the scope of this work. Instead, we focus on one model where we search for dust dynamics solutions that capture general features of dust trajectories.  
Here we use the case of a faster rotating star, $\Omega_\star=0.2\Omega_{\mathrm br}$ with a smaller stellar magnetic field of 750 Gauss than used in \cite{Mishra}, which for $\alpha_{\mathrm v}=\alpha_{\mathrm m}=1$ shows intermittent backflow. This indicates that the critical value of P$_{\mathrm m}\sim 0.6$ for the presence of backflow in the mid-plane of the disc depends on the stellar rotation rate\footnote{In the conclusions in \cite{Mishra} such cases were relegated to future work.}. We choose this case because of density distribution atop the disc showing vertical variations, which can help to better understand behaviour of the dust atop the disc. A couple of snapshots of gas velocity and density from the chosen model are shown in Fig.~\ref{fig:MHD}. The model parameters are listed in Table~\ref{tbl:parameters}.

In \cite{ckp23} analytical results on the MHD thin discs were directly compared with the numerical simulations. One of the main outcomes was that magnetic field penetrating the thin disc does not affect the solutions inside the middle part of the disc - they remain virtually the same as in the non-magnetic cases. Close to the inner disc rim and outer boundary of the computational box, numerical solutions are inevitably defined by the numerical setup. Star-disc magnetospheric interaction is most violent atop the disc and is governed by the geometry of the reconnection layer above the disc, where the stellar field meets with the disc field. In some cases this layer is pushed radially outwards, beyond the outer boundary of the computational box, but in most cases a part of it remains inside the box \citep[see the four types of solutions in Fig.~1 in][]{CemBrun23}. The position of the current sheet along this layer strongly influences the atmosphere above the disc, the ``disc corona''. This is where, as a back-reaction of the outer boundary, in our chosen case a column of density occurs, which we employ to illustrate the effects of variations in the disc corona on the material deposited at the disc surface. The exact positioning of such a column could vary in the cases with different parameters, but the underlying processes remain the same.

\begin{table}
 \caption{T~Tau stellar properties and parameters of the disc wind model used in this study.}
 \label{tbl:parameters}
 \begin{tabular}{ccccccc}
  \hline
 $T_*$ & $M_*$     & $R_*$     & $L_*$     & $\gamma$ & $R^{\mathrm{thick}}_{\mathrm{ in}}$\\
 K     & $M_\odot$ & $R_\odot$ & $L_\odot$ &  & $R_*$  \\
  \hline
 3700 & 0.5 & 2 & 0.67 & 0.90 & 6.38\\
  \hline
  \hline
 $\alpha_{\mathrm v}$ & $\alpha_{\mathrm m}$ & $B_\star$ & $P_\star$ & $\dot{M}_{\mathrm acc}$ & $L_{\mathrm acc}$\\
           &            &  kG & d& 10$^{-9}M_\odot\,y^{-1}$ & $L_\odot$ \\
  \hline
 1 & 1 & 0.75 & 2.32 & 11.4 & 0.073\\
  \hline
 \end{tabular}
\end{table}

\subsection{Theory}
\label{Sec:TTau_trajectory}

The approach how to combine the output from MHD modelling and the physics of dust dynamics is described in our previous paper \citetalias{Vinkovic21}. There are two key updates in our current work. The first is introduction of time evolution of the gas density and velocity. This time dependence is the complete output from the MHD model, while previously we simplified our work by taking a static snapshot of the MHD disc wind. The second update is introduction of the non-radial radiation pressure force originating from the diffuse thermal radiation emitted by the hottest dust at the inner dusty disc edge. This addition is the same as the one described in the previous section on Herbig Ae stars. 

The gas density $\rho_{gas}(\bmath{r},t)$ now depends on a position in space ($\bmath{r}$) an time ($t$). This time dependence is then propagated to various other variables in the numerical model that we use:
\begin{itemize}
    \item[--] the optical depth $\tau(\bmath{r},t)$ integrates over dust density, which we correlate with the gas density $\rho_{gas}(\bmath{r},t)$,
    \item[--] the vector $\bmath{q}_{ext}(\bmath{r},t)$ \citepalias[from equation 19 in][]{Vinkovic21} because it depends on $\tau(\bmath{r},t)$,
    \item[--] the "strength" of the radiation pressure force vector $\bmath{\beta}(\bmath{r},t)$ (that we already described in equation \ref{EqBeta}, but now we will expand it further) because it depends on $\bmath{q}_{ext}(\bmath{r},t)$,
    \item[--] the gas and dust temperature $\bmath{T}(\bmath{r},t)$ because it depends on the optical depth structure \citepalias[equation 26 in][]{Vinkovic21},
    \item[--] the dimensionless parameter $\mu(\bmath{r},t)$ \citepalias[equation 30 in][]{Vinkovic21} because it depends on $\rho_{gas}(\bmath{r},t)$ and $\bmath{T}(\bmath{r},t)$.
\end{itemize}

The strength of radiation pressure vector $\bmath{\beta}(\varrho,z,t)$ in the cylindrical coordinate system ($\varrho$,$z$) is technically still the same as in equation \ref{EqBeta}, except that we need to readjust the scaled radiation pressure vector $\bmath{\mathfrak{f}}(\varrho,z,t)$ to better represent the stellar radiation from T Tau stars. The readjustment combines the old stellar radiation pressure from equation 21 in \citetalias{Vinkovic21} and the new non-radial component $\mathfrak{f_\theta}$, which gives
\begin{equation}\label{EqBetaTTau}
\begin{split}
\bmath{\beta}(\varrho,z,t)=0.19 \left(\frac{L_*}{L_\odot}\right)\left(\frac{M_\odot}{M_*}\right)
\left(\frac{3000 \, \text{kg}\, \text{m}^{-3}}{\rho_{\mathrm{grain}}}\right)\left(\frac{\mu \text{m}}{a}\right) \\
\cdot\left(\bmath{q}_{\mathrm{ext}} + |\bmath{q}_{\mathrm{ext}}|\mathfrak{f_\theta}\bmath{\hat{\theta}} \right),
\end{split}
\end{equation}
where $\mathfrak{f_\theta}$ is the same as before (shown in Fig.~\ref{fig:HAeFig1}), just scaled to the T~Tau disc size.

\begin{figure}
 \includegraphics[width=\columnwidth]{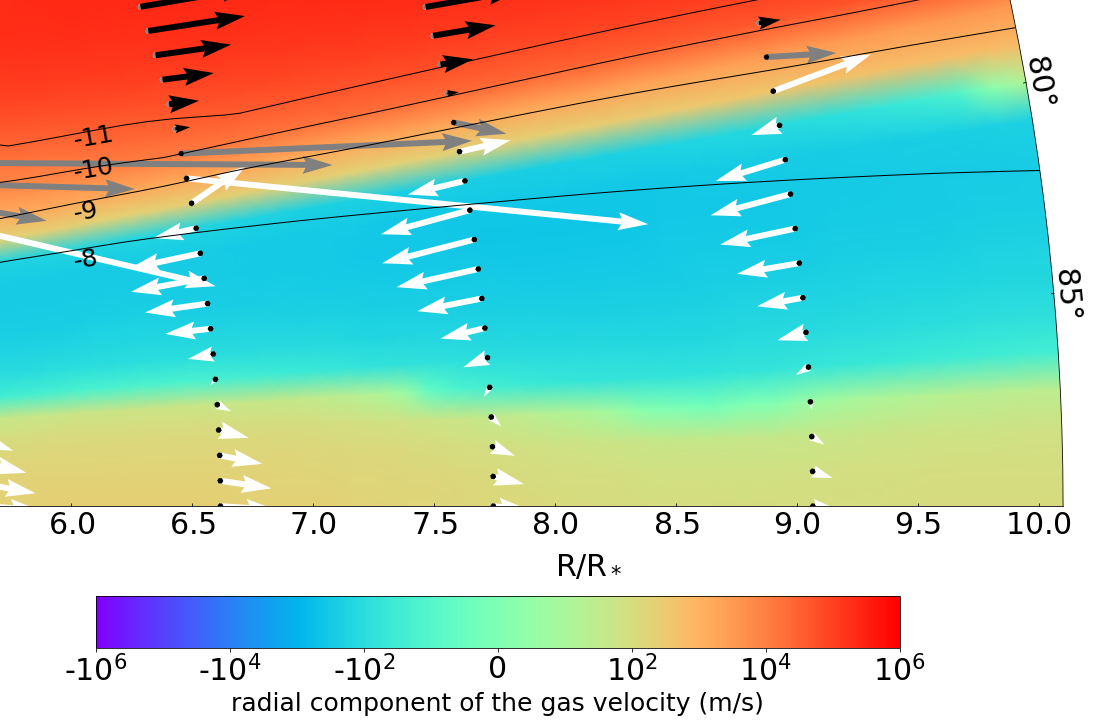}
 \caption{Zoom on the region where the inner edge of dusty disc is located in our simulations. The snapshot of MHD model is the same as the bottom row in Fig. \ref{fig:MHD}, with the same colour and arrow representation of gas velocity. The gas density is shown as contours, where numbers are the decimal exponent of gas density in kg~m$^{-3}$.}
 \label{fig:MHDinnerdge}
\end{figure}

\begin{figure}
 \includegraphics[width=\columnwidth]{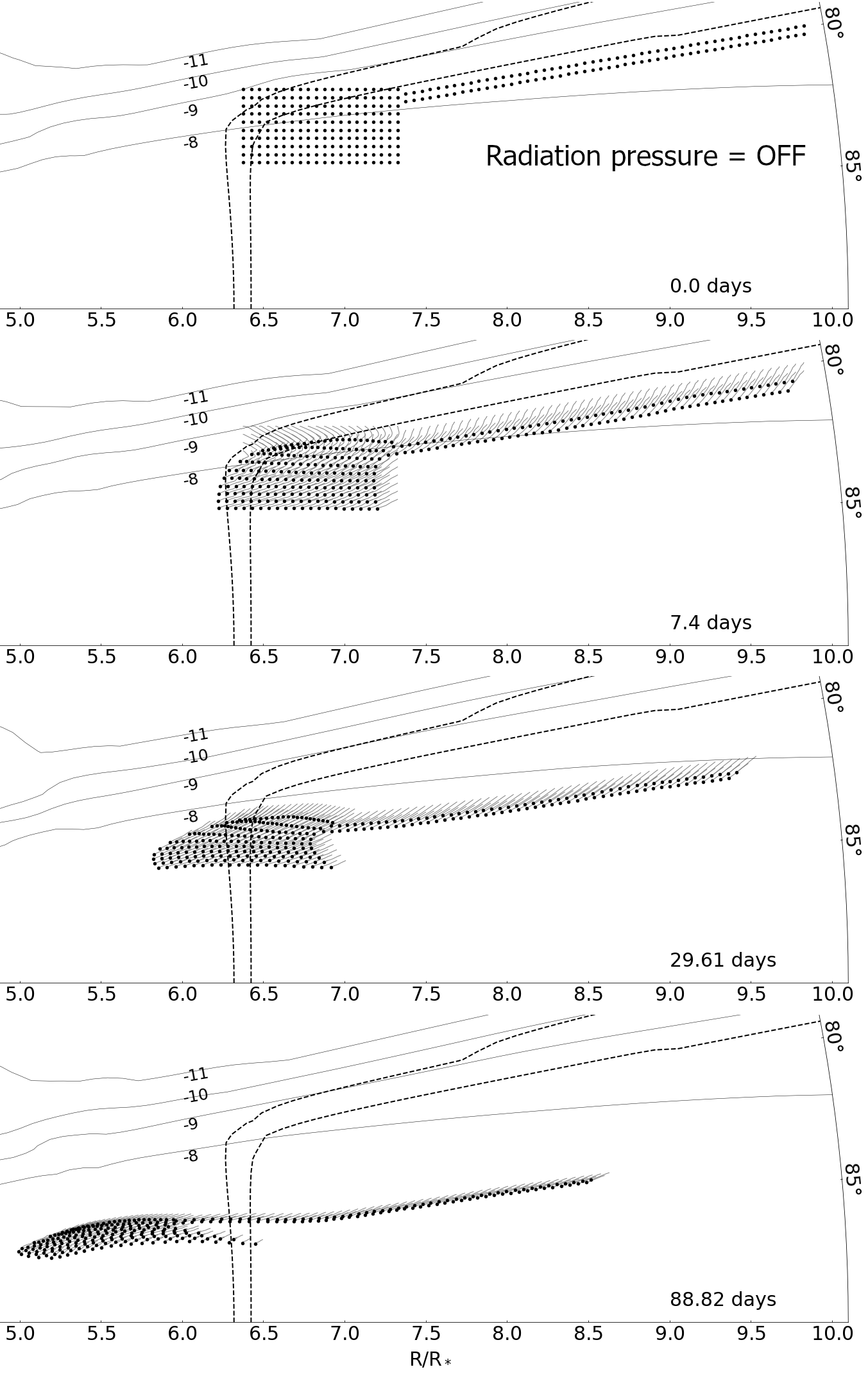}
 \caption{Dynamics of dust grains at the inner edge of optically thick dusty disc around a T~Tau star when we ignore the radiation pressure force. The dots are positions of dust grains at different times. The simulation starts with the grains distributed as shown in the top panel, marked as zero days. The other panels show grains' positions at later times, as indicated by days since the start. Each point has a thin grey solid line tail indicating its path during the previous 7.4 days. For comparison, the orbital Keplerian period is 0.464$(R/R_*)^{3/2}$~days, which gives 7.47~days at the optically thick dusty disc radius $R^{\mathrm{thick}}_{\mathrm{in}}$. Grains are 4$\mu$m in size. The dashed lines show the position of dusty disc surface dictated by the radial optical depth parameter $\varepsilon$ described by equation 16 in \protect\citetalias{Vinkovic21}, where the dashed lines are $\varepsilon=e^{-0.1}$ and $\varepsilon=e^{-1}$. The gas density is shown as solid lines contours, where numbers are the decimal exponent of gas density in kg~m$^{-3}$. Animation showing the grains' movement is available in the supplementary materials as \texttt{Figure\_7.mp4} (see Appendix \ref{AppendixAnima}).}
 \label{fig:Inner4micNOradPress}
\end{figure}

\begin{figure}
 \includegraphics[width=\columnwidth]{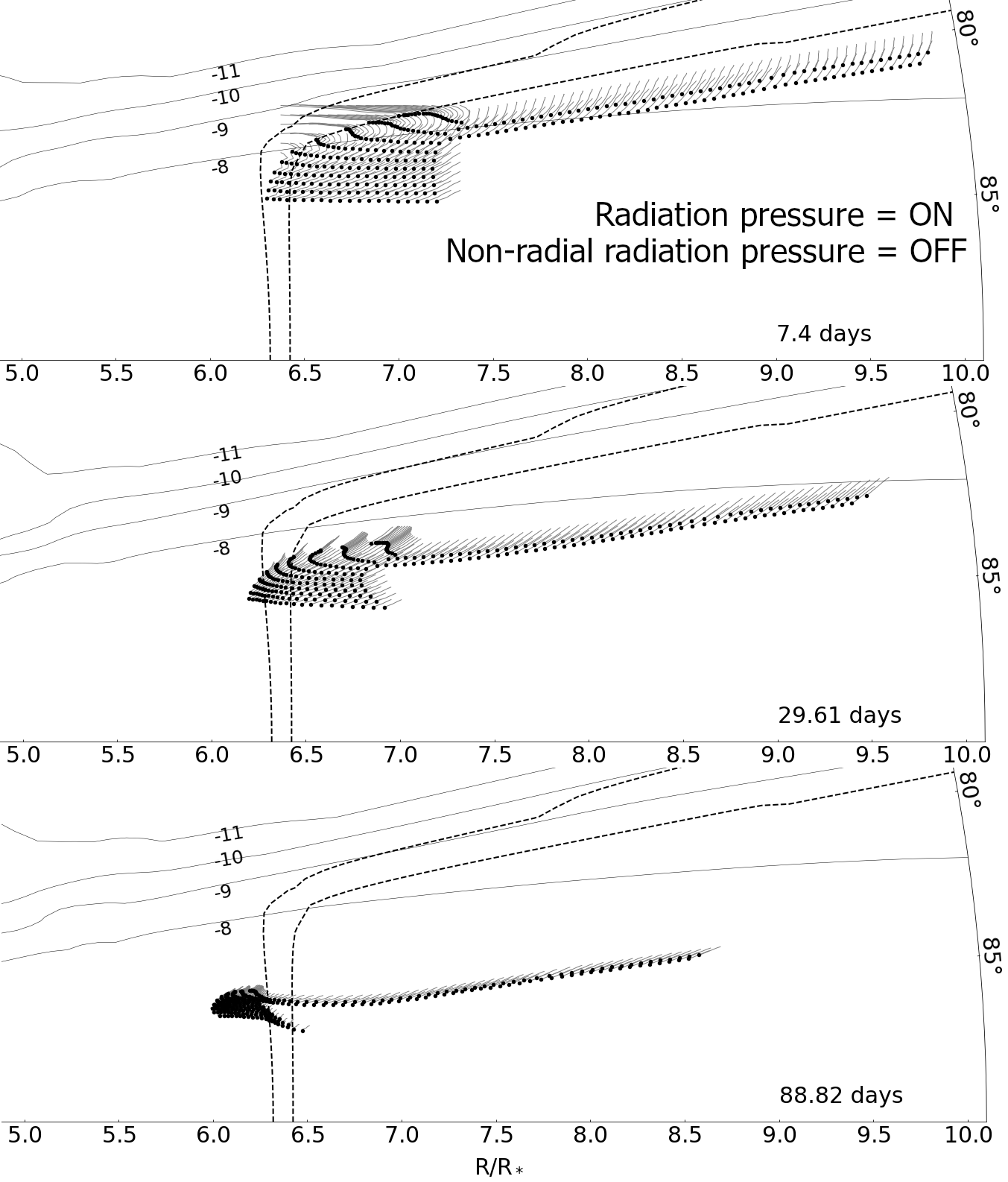}
 \caption{The same as Fig.~\ref{fig:Inner4micNOradPress}, but with included radiation pressure. However, the non-radial radiation pressure, emanating from the hot dust, is ignored. Animation is available in the supplementary materials as \texttt{Figure\_8.mp4} (see Appendix \ref{AppendixAnima}).}
 \label{fig:Inner4micNonRadialOff}
\end{figure}

The stellar parameters used for T~Tau stars are shown in Table~\ref{tbl:parameters}. They result in the inner dusty disc radius of $R^{\mathrm thick}_{\mathrm in}$=6.38$R_*$. This is the radius where the disc becomes optically thick, while optically thin dust made of big grains can move closer to the star \citep[for futher discussion see][]{Vinkovic06,Kama09,Vinkovic12,Flock16}. We improved slightly equation 7 from \citetalias{Vinkovic21} to better describe\footnote{The updated version is:\\ $\log(\xi^{\mathrm{thin}}(r))=\log(\xi^{\mathrm{thick}}) + \left(5 {\log(r/R^{\mathrm{thin}}_{\mathrm{in}}) } / { \log(R^{\mathrm{thick}}_{\mathrm{in}}/R^{\mathrm{thin}}_{\mathrm{in}})} -7\right)$} this optically thin part of the inner disc edge in accordance with the solution by \cite{Flock16}.

The snapshots of MHD simulation are saved periodically with some time step $\delta t$. In our case this step is $\delta t=0.1P_*$, where $P_*=2.32$~d is the stellar period, which is small enough to capture a smooth transition of gas velocity and density from one step to another. The problem now is how to extract values at moments that exist between two consecutive snapshots. We use a simple linear interpolation for that purpose. 

If the time since the beginning of simulation is $t$, and it falls between snapshots $n$ at time $t_n=n\delta t$ and $n+1$ at time $t_{n+1}=(n+1)\delta t$, then an interpolated variable $K(\varrho,z,t)$ is
\begin{equation}\label{interpolation}
K(\varrho,z,t) = K(\varrho,z,t_n)(1-\Delta t)+K(\varrho,z,t_{n+1})\Delta t,
\end{equation}
where $0\leq \Delta t<1$ and
\begin{equation}\label{interpolationDeltaT}
\Delta t =\frac{t-n\delta t}{\delta t}.
\end{equation}

Now we have all the updates required for the equations of dust motion \citepalias[equations 31-33 in][]{Vinkovic21} in the cylindrical coordinate system $\bmath{r}=(\varrho,\varphi,z)$ 
\begin{equation}\label{eq:ForceRhoTtau}
\ddot{\varrho}=\varrho \dot{\varphi}^2 - \frac{\varrho}{(\varrho^2+z^2)^{3/2}} - \mu(t) (\dot{\varrho}-\text{v}_{\mathrm{gas},\varrho}(t))+\frac{\beta_\varrho(t)}{\varrho^2+z^2}
\end{equation}
\begin{equation}\label{eq:ForcePhiTtau}
\ddot{\varphi}=-2\frac{\dot{\varrho}}{\varrho}\dot{\varphi}-\mu(t) \left(\dot{\varphi}-\frac{\text{v}_{\mathrm{gas},\varphi}(t)}{\varrho}\right)
\end{equation}
\begin{equation}\label{eq:ForceZTtau}
\ddot{z}= - \frac{z}{(\varrho^2+z^2)^{3/2}} - \mu(t) (\dot{z}-\text{v}_{\mathrm{gas},z}(t))+\frac{\beta_z(t)}{\varrho^2+z^2}
\end{equation}
where the radiation pressure vector from equation \ref{EqBetaTTau} is axially symmetric $\bmath{\beta}(t)=(\beta_\varrho(t),0,\beta_z(t))$, and the gas velocity is $\bmath{V}_{\mathrm{gas}}(t)=(\text{v}_{\mathrm{gas},\varrho}(t),\text{v}_{\mathrm{gas},\varphi}(t),\text{v}_{\mathrm{gas},z}(t))$. 

The pseudo-code describing the sequence of computational steps is the same as in \citetalias{Vinkovic21}, except that now we must extract the values of time-dependent variables using the interpolation equation \ref{interpolation}.

\begin{figure}
 \includegraphics[width=\columnwidth]{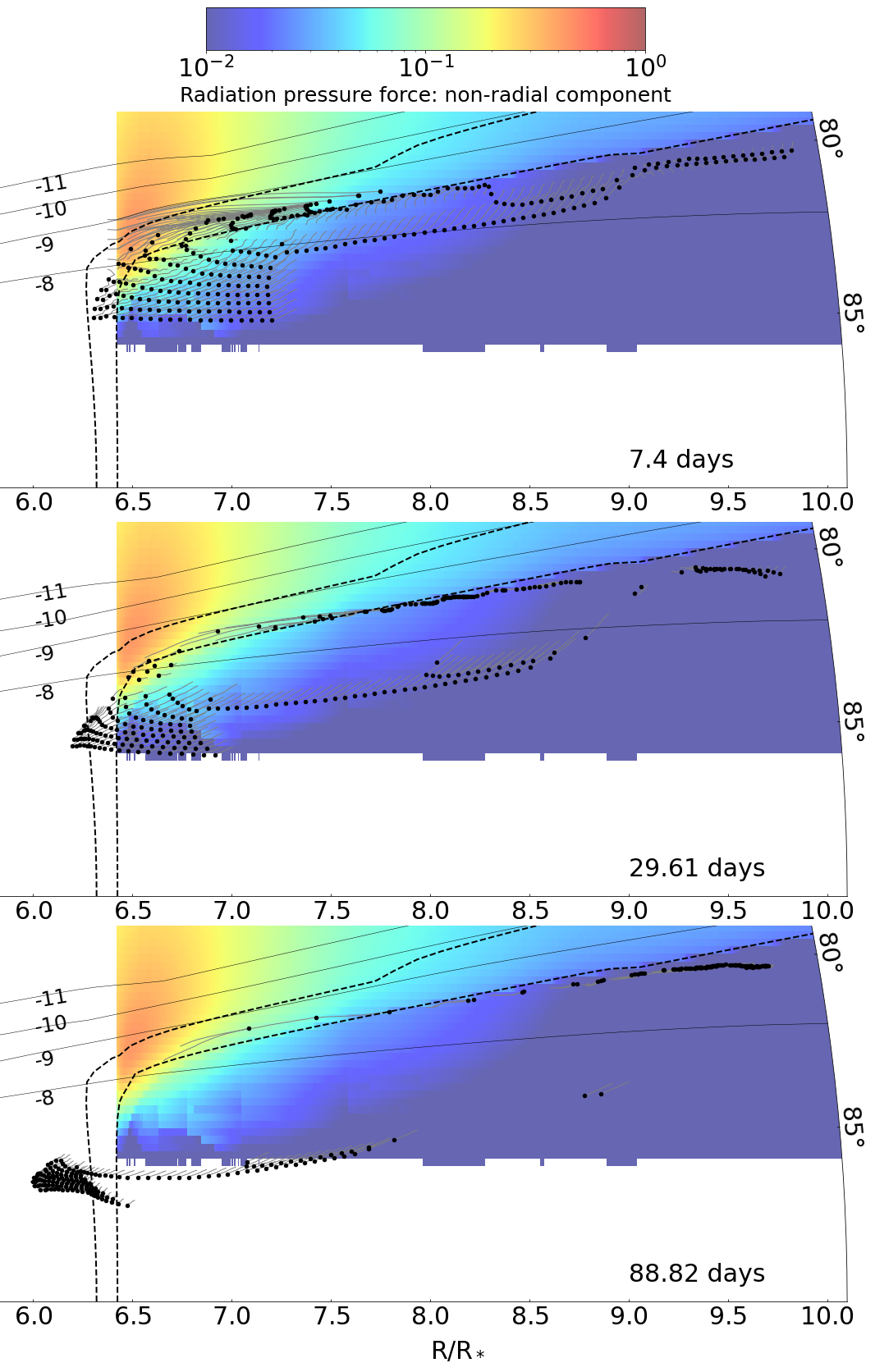}
 \caption{The same as Fig.~\ref{fig:Inner4micNOradPress}, but with the complete description of radiation pressure force included into the dust dynamics. The colour map shows the non-radial component of radiation pressure (see equation \ref{EqBetaTTau} and Fig.~\ref{fig:HAeFig1}). Animation is available in the supplementary materials as \texttt{Figure\_9.mp4}  (see Appendix \ref{AppendixAnima}).}
 \label{fig:Inner4mic}
\end{figure}

\subsection{Inner edge}

We investigate the behaviour of dust at the inner dusty disc edge in three steps. First, we completely ignore the radiation pressure force. Then we repeat the same simulation with only the stellar component of radiation pressure turned on. Finally, we also introduce the non-radial contribution to the radiation pressure force. This approach enables us to inspect the importance of the full description of radiation pressure in the dust dynamics. 

Fig.~\ref{fig:MHDinnerdge} shows the gas density and velocity distribution in this part of the disc based on the MHD model. Notice how the gas changes its radial velocity direction from outflow above the dense disc to inflow in the upper regions of the disc and then back to outflow in the midplane. This midplane backflow was recently shown to be a physical feature and not a numerical artefact (see section \ref{SubSec_disccorona}). Another important feature at the disc surface, where the radial velocity flips its direction, is a fast drop in the gas density accompanied by a dramatic rise in the gas speed. This interplay between the magnitude of gas density and velocity is interesting because the gas drag is proportional to both. Hence, the drag might stay strong above the disc as long as the rise in the gas speed compensates for the drop in the gas density. 

We use olivine grains of 2~$\mu$m in radius, with the optical properties taken from \cite{Dorschner95}. Grains in the micron range in size display a similar dynamical behaviour. Submicron grains are too small to be significantly affected by the non-radial radiation pressure, which leads to the dust dynamics described in \protect\citetalias{Vinkovic21} where only the stellar radiation pressure is explored. Larger grains experience drops in both the gas drag and the radiation pressure force, which makes it out of the scope of our paper. The grain bulk density is 1000~kg~m$^{-3}$, which is more similar to porous fluffy grains.

We start with the model where the radiation pressure component in equations \ref{eq:ForceRhoTtau}-\ref{eq:ForceZTtau} is not included (i.e., $\bmath{\beta}(t)=0$).  Fig.~\ref{fig:Inner4micNOradPress} shows how the dust moves under such conditions. Initially we distribute grains in the most interesting parts of the inner disc and then observe how they move. Grains are sinking deeper into the disc and are being dragged toward the star by the gas. The grains pass through the dusty disc edge without interruption because these grains are big enough to survive sublimation at that distance and there is no radiation pressure force to slow them down. 

A significant change happens when we introduce the stellar radiation pressure in Fig.~\ref{fig:Inner4micNonRadialOff}. The drift of dust particles toward the star is stopped when they emerge from the optically thick interior and become exposed to the direct stellar radiation. The dust accumulates at $R^{\mathrm{thick}}_{\mathrm{in}}$ and settles downward. We are still ignoring the non-radial contribution to the radiation pressure, which makes this solution equivalent to the results described in \protect\citetalias{Vinkovic21}. They show how the midplane gas backflow can capture such dust and transport it back into the disc. 

Finally, we utilise the full description of the radiation pressure, where we include the non-radial radiation pressure force component $\mathfrak{f_\theta}$ in equation \ref{EqBetaTTau}. For this we exploit the intrinsic scaling property of the equations of dust radiative transfer according to which luminosity and linear dimensions are irrelevant for solving the equations \citep{IE97}. Luminosity is important only when we need to translate dimensionless radiative transfer solutions to physical units, where physical dimensions scale with the square root of luminosity. Hence, we import the same solution for $\mathfrak{f_\theta}$ as in Fig.~\ref{fig:HAeFig1} \citep[originally from][]{Vinkovic09}, just scaled by $(L_\star+L_{\mathrm{acc}})^{1/2}$ for T~Tau star.

Fig.~\ref{fig:Inner4mic} shows how the non-radial component $\mathfrak{f_\theta}$ is positioned to fit the dusty disc structure around our T~Tau star. The dust is initially released from the same positions as before, but their trajectories display now two types of scenarios where the grains end up. The grains that are deep enough within the optically thick disc to avoid the influence of $\mathfrak{f_\theta}$ tend to follow paths similar to the previous solution in Fig.~\ref{fig:Inner4micNonRadialOff}. These grains accumulate at the inner disc edge where they settle toward the midplane. The grains exposed to the non-radial radiation pressure detach from this type of paths and move first upward and then outward along the disc surface. They are deposited at some point further back in the disc. This creates conditions where dust experiences a convective motion not triggered by gas disc turbulence but driven by the non-radial component of radiation pressure force.

\begin{figure}
 \includegraphics[width=\columnwidth]{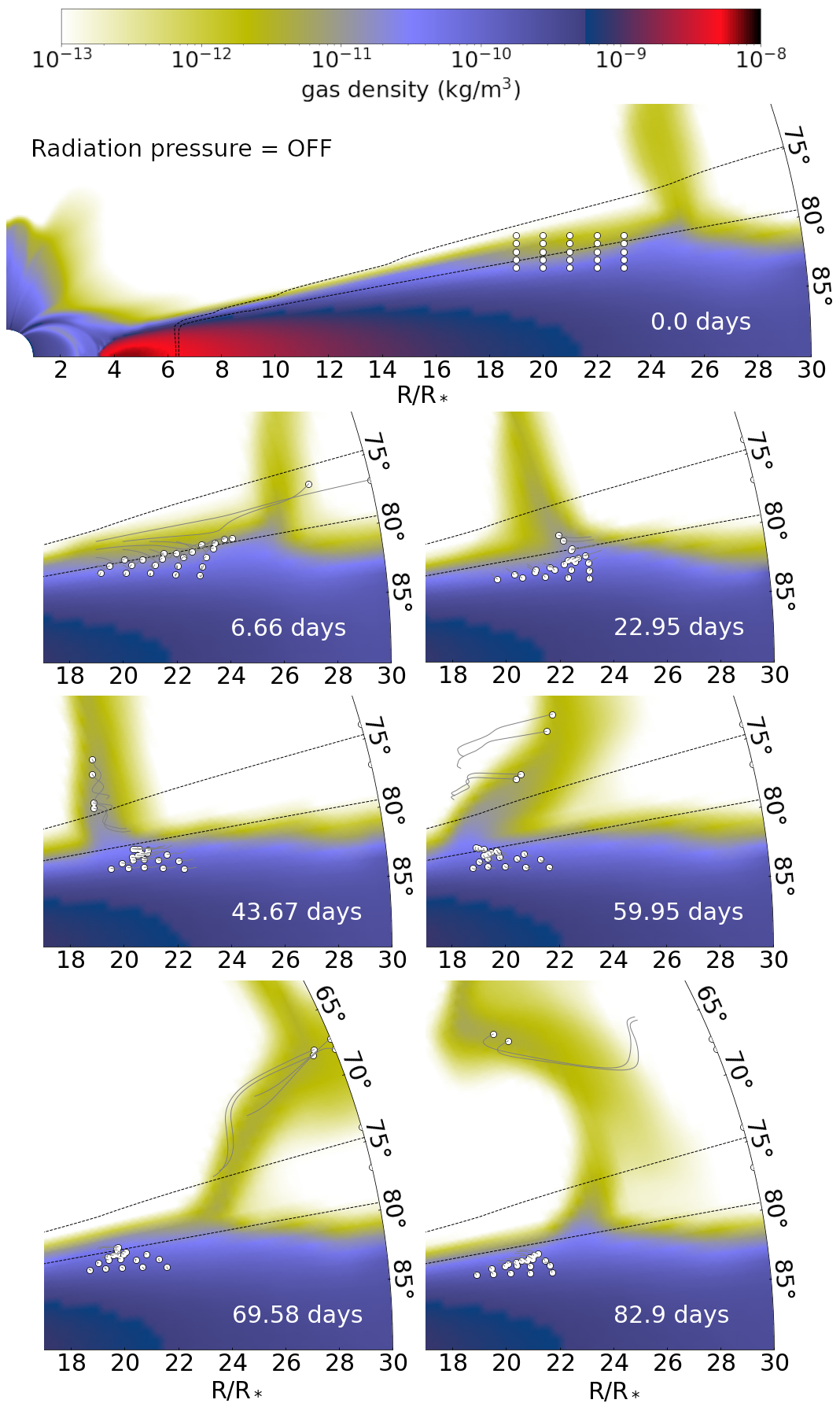}
 \caption{Interaction between the MHD gas wind and the dusty disc when the radiation pressure force is ignored. The dust particles traced by simulation are initially positioned as shown in the top panel, marked as zero days. The other panels show grains' positions at later times, as indicated by days since the start. Each point has a thin grey solid line tail indicating its path during the previous 7.4 days. For comparison, the orbital Keplerian period is 0.464$(R/R_*)^{3/2}$~days, which gives 41.5~days at $R=20R_*$. Grains are 1$\mu$m in size and  1000~kg~m$^{-3}$ in bulk density. The dashed lines show the position of dusty disc surface dictated by the radial optical depth parameter $\varepsilon$ described by equation 16 in \protect\citetalias{Vinkovic21}, where the dashed lines are $\varepsilon=e^{-0.1}$ and $\varepsilon=e^{-1}$. The colour map shows the gas density. Animation is available in the supplementary materials as  \texttt{Figure\_10.mp4} (see Appendix \ref{AppendixAnima}).}
 \label{fig:05noPress}
\end{figure}

\begin{figure}
 \includegraphics[width=\columnwidth]{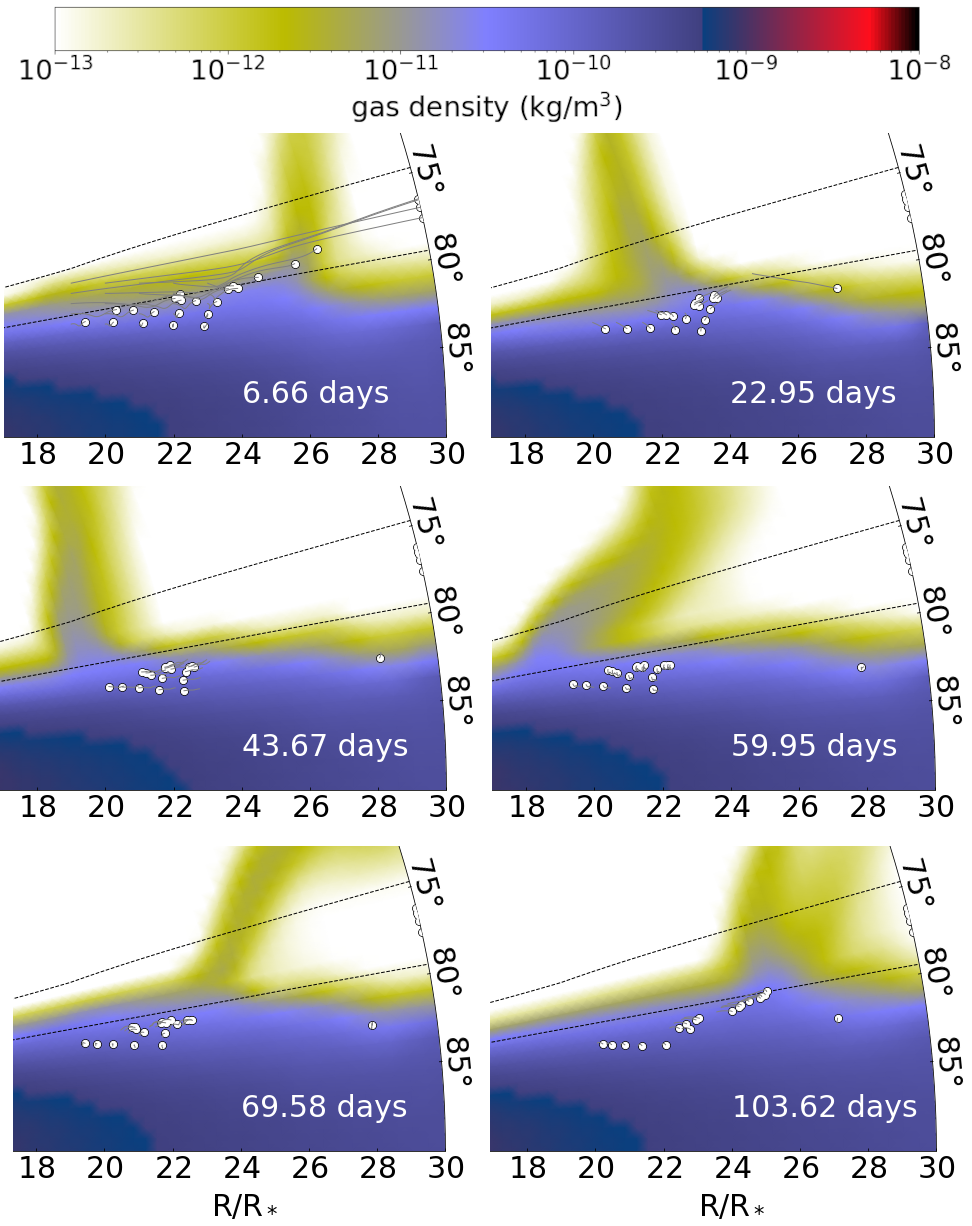}
 \caption{The same as Fig.~\ref{fig:05noPress}, but with included radiation pressure. Animation is available in the supplementary materials as \texttt{Figure\_11.mp4} (see Appendix \ref{AppendixAnima}).}
 \label{fig:05withPress}
\end{figure}

\begin{figure}
 \includegraphics[width=\columnwidth]{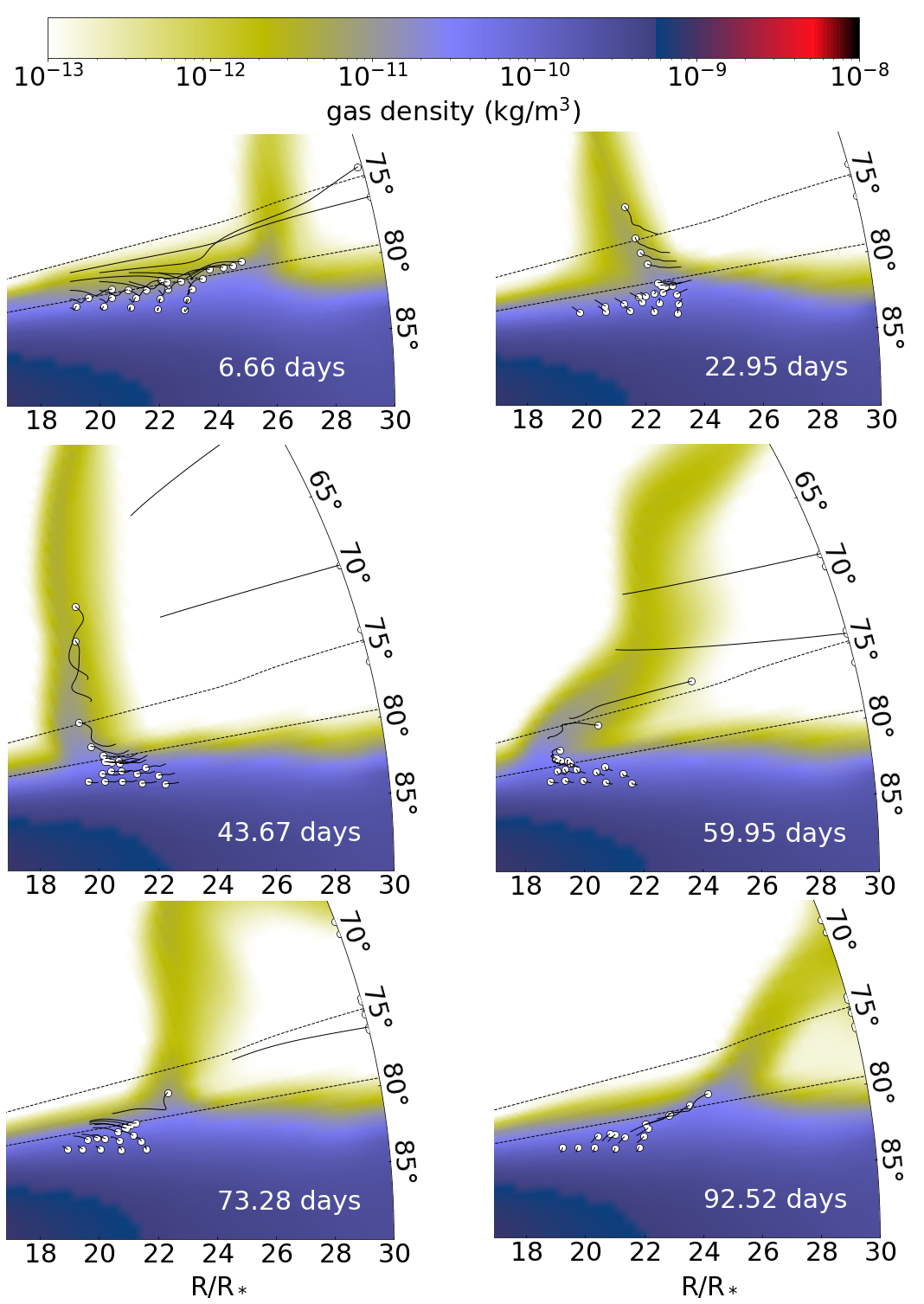}
 \caption{The same as Fig.~\ref{fig:05noPress}, but with included radiation pressure and for grains of 0.2$\mu$m in size and 3000~kg~m$^{-3}$ in bulk density. Animation is available in the supplementary materials as \texttt{Figure\_12.mp4} (see Appendix \ref{AppendixAnima}). }
 \label{fig:01withPress}
\end{figure}

\begin{figure}
 \includegraphics[width=\columnwidth]{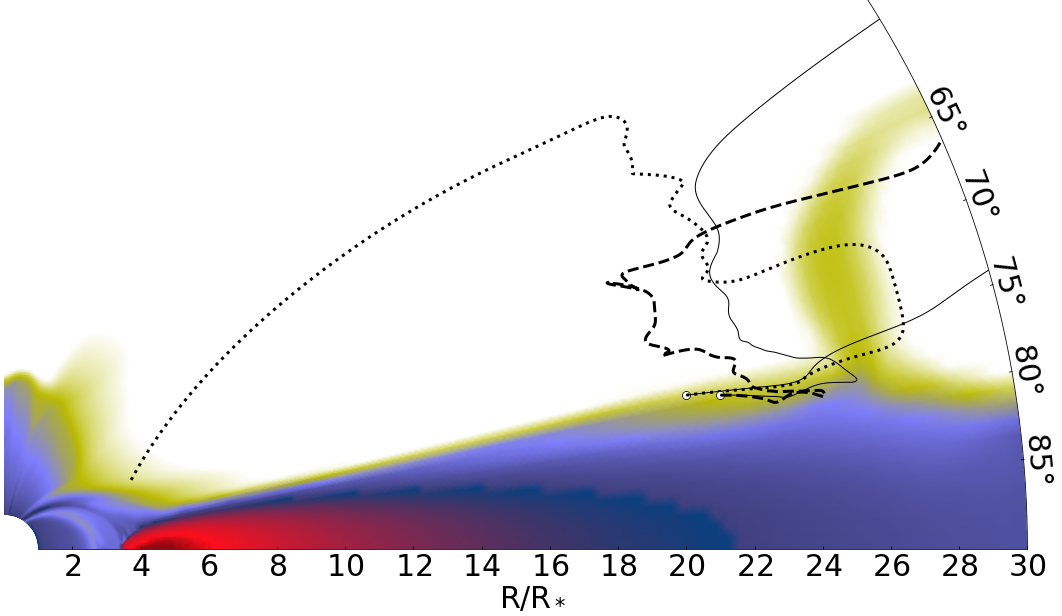}
 \caption{Comparison between trajectories for two grains of 0.2$\mu$m in size computed with (solid lines, the same as Fig.~\ref{fig:01withPress}) and without (dashed and dotted lines) radiation pressure. This example emphasises the most dramatic impact of radiation pressure, where it does not allow dust to move inward. Instead, the radiation pressure force pushes the dust outward. Animation showing the grains' movement without radiation pressure is available in the supplementary materials as \texttt{Figure\_13.mp4} (see Appendix \ref{AppendixAnima}). Compare this animation with the results in Fig.~\ref{fig:01withPress} where the radiation pressure is not ignored.}
 \label{fig:01NoPress}
\end{figure}

\subsection{Disc surface away from the inner edge}
\label{SubSec_currentsheet}

At some distance outward of the inner disc edge we expect a current sheet to appear in the disc wind. Even though our model is not a precise reconstruction of this phenomenon, we can investigate the dust dynamics behaviour at that part of the disc and possibly extract some robust general conclusions. In our MHD model the current sheet appears approximately at 20-26 stellar radii form the star. It is constantly in motion, disturbing the disc surface. 

We position dust particles in the disc surface and run dust dynamics simulation with and without the radiation pressure included. In these disc regions we do not expect a non-radial radiation pressure because both the dust temperature and the dust emissivity are too low. The top panel in Fig.~\ref{fig:05noPress} shows the initial positions of dust grains used in the example we show here. 

Unlike in the inner dusty disc edge, the disc surface at distances from the star considered here is populated with small grains. What interests us are grains that are most affected by the stellar radiation pressure. In Fig.~\ref{fig:05noPress} we show the trajectories of grains 1$\mu$m in size and 1000~kg~m$^{-3}$ in bulk density, but without radiation pressure included into the dust dynamics equations. Even though grains tend to get dragged deeper into the disc by the dense disc gas, some dust grains are captured by the density column above the disc. Once they become lifted above the disc, they are at the mercy of chaotic gas velocity changes. Their paths become highly complicated and their destination unpredictable. Also, they can be lifted quite high, an entire geometrical disc thickness or more above the disc, as long as they are supported by the gas within the density column. Eventually they escape this trap and re-enter the disc either inward or outward. 

The situation changes dramatically when we include the radiation pressure force. Fig.~\ref{fig:05withPress} shows the paths of the same dust when they experience the stellar radiation pressure. The outcomes are now much simpler as the grains lifted out of the disc are quickly pushed outward. They exit our computational domain, but we can expect them to re-enter the disc at some larger distance. The stark contrast between these two solutions tells us that such grains cannot be properly explored if we ignore the radiation pressure force.

An interesting question now is whether dust grains can reach considerable heights above the disc or not? In Fig.~\ref{fig:01withPress} we show an example with grains of 0.2$\mu$m in size and 3000~kg~m$^{-3}$ bulk density. The grains are initially located at the same positions as before. The figure shows dust paths with the radiation pressure included and this time grains are lifted high above the disc. The gas drag is strong enough within the density column to trap the dust, but the radiation pressure is again crucial for understanding the final destiny of such grains. All of them eventually end up blown away into the outer disc regions. When we compare this solution with the one without the radiation pressure, we encounter the same critical distinction as before. Fig.~\ref{fig:01NoPress} shows this comparison on the paths of two grains that exhibit two major differences in their behaviour when the radiation pressure is not included: they spend significantly longer times trapped above the disc, and they can end up being ejected inward. Thus, the radiation pressure force should not be ignored. It is necessary because it ejects dust outward and shortens the time dust spends above the disc. 

\section{Discussion}
\label{SecDiscuss}

The diversity of our dust dynamics solutions illustrates difficulties that we face in reconstructing the physics of inner regions of dusty discs. The initial hope that simple steady-state geometrical structures \citep{Review2010} would suffice is no longer valid. Advancements in observational techniques reveal diversity of disc structures and the need for more advanced modelling \citep{GRAVITY_HAe,GRAVITY_TTau,Review2023}. Observations of inner discs are affected by the light originating from the surface dust exposed to the direct stellar illumination. This puts radiation pressure into the centre of interest when the disc structure is deduced from such observations. 

Our results testify to the importance of radiation pressure force in this context. We show how it erodes the upper parts of the inner disc wall, which sets upper limits on the surface responsible for the near infrared emission \citep{Vinkovic14}. The dust dynamics of this erosion is enhanced by the non-radial radiation pressure originating from the diffuse infrared dust emission \citep{Vinkovic09}. 

In Herbig Ae stars these effects have a dramatic impact on the disc structure and dust dynamics. A large stellar luminosity drives a strong radiation pressure force on the disc surface dust\footnote{Notice that \cite{Vinkovic09} underestimated the strength of radiation pressure force compared to our work.}, which results in a sharp transition between the zone dominated by the gas drag and the zone where radiation pressure clears the dust away. We used a simple Keplerian rotation for the disc gas. A more realistic approach could describe non-trivial gas dynamics, which would affect the location of transition layer and lead to transiently enhanced dust outflows. Our results suggest that this also holds in T Tau stars, even though their luminosity is smaller and, therefore, the radiation pressure effects are less dramatic. 

Removing dust from the disc surface leads to the problem of surface dust replenishment. While the surface of inner disc edge can be replenished by simple inward gas accretion, the rest of the disc surface requires other mechanisms. Aerodynamically well coupled dust grains can be stirred upward by the vertical upflows accompanying MHD or photoevaporation disc winds \citep[e.g.,][]{Owen2011,Bans, Flock17, Riols,Giacalone,Franz}. Our results emphasise the importance of radiation pressure force for the dust launching mechanism in such disc winds and this contribution should not be ignored. 

We can also expect the radiation pressure effects to persist even in the case of different dusty disc geometries \citep[due to more realistic treatment of dust sublimation, chemistry, grain size distribution, and gas dynamics – e.g., ][]{IsellaNatta,Kama09, Flock17}, since the key prerequisite is only the stellar illumination. If the dust temperatures reach peak emission in the near infrared (e.g., sublimation of silicates), then the non-radial contribution to the radiation pressure force will also exist regardless of the disc geometry. We show how this non-radial force induces vertical mixing of dust at the inner disc edge. 

A more advanced approach would also incorporate magnetically driven disc turbulence and ambipolar diffusion that can transport dust back into the disc surface \citep{Flock17,Riols}. This can expose dust to multiple reheating episodes. The role of dust charging also remains an open question, as we encounter dusty plasma and dust ionization in the disc surface \citep{Ivlev}. Charged dust would interact with the magnetic field, which might significantly impact the dynamics of this dust.

As we already explained in section \ref{SubSec_disccorona}, the MHD model used in our study is just an example that illustrates dust dynamics possibilities. We also argued that the midplane backflow seen in the disc gas velocity field is real. The ability of radiation pressure force to trap dust at the inner disc wall is quite interesting as it can interact with the midplane backflow. This might create a mechanism of transporting thermally processed dust further back into the disc midplane and contribute to the planet formation process. Hence, it would be interesting to study this particular feature of the dust and gas dynamics into more detail. 

The immediate vicinity of stars in T Tau objects is also interesting because this is the region where dust clouds produce episodic or quasiperiodic dimming of stellar light in variable stars known as "dippers" \citep{dippers2017,dippers2021}. Since the dust dynamics is the key part of their modelling, our paper shows that radiation pressure force should not be ignored in these models. We also know from the radiative transfer principles that large dust grains survive the closest to the star 
\citep{Vinkovic06,Kama09}. A more advanced version of our model should also follow big dust grains that survive all the way to the corotation radius and enter the accretion funnel flow \citepalias[see also][]{Vinkovic21}. Dust evaporation takes time \citep{dustdiffusion} and it is influenced by the local gas pressure and chemistry \citep{Duschl}, especially in big grains, which might be just long enough to accumulate dust close to the star and dim its light \citep{magnetosphere}. Indeed, big grains have been identified as the source of dimming in some objects \citep{Sitko2023}. 

A more common source of dimming is probably dusty irregularities connected to the turbulent nature of discs and their geometrical distortions. Outbursts in the stellar activity can also temporarily increase the radiation pressure force over the inner dusty disc surface and push the warm surface dust outward \citep{EXLup}. Interestingly enough, transient dust outflows might happen also because of the disc instabilities induced by the radiation pressure at the inner dusty disc edge. We show here how dust in that zone is pushed inward by the gas accretion, but at the same time grains experience outward radiation pressure force. If enough dust accumulates in this force limbo, optical depth effects on the disc gas locally perturb the disc stability and lead to clumping \citep{instability}. Such clumps might be responsible for occasional dimming of the star or transient dust outflows. 

Another interesting feature addressed in section \ref{SubSec_currentsheet} is interplay between the MHD disc wind's current sheet dynamics and the radiation pressure force on dust trapped within the current sheet. Our results can only indicate a qualitative general behaviour of dust particles in this case, where the radiation pressure is responsible for a fast outward ejection of dust while the surface gas flow pulls particles out of the disc. More advanced MHD models are needed to produce possible comparisons with observations, such as multiwavelength photometric variability caused by the complicated dust and gas dynamics in this part of the disc surface.

Dust grains close to the inner dusty disc edge are hot and sticky \citep{sticky}, which leads to the creation of porous fluffy aggregates. This enhances the radiation pressure force effect \citep{fluffy2015} as well as the gas drag force \citep{porousdrag}. We address this only marginally by using a smaller bulk density of big grains. Further improvements are possible on the description of these forces on big porous grains, which would enhance the dust outflow in our model. 

In situ measurements of cometary dust also reveal that dust grains are actually porous aggregates of individual grains of about 1~$\mu$m in size \citep{fluffy2016}. Analysis of ablation and fragmentation of meteors during flight also indicates low density and fluffiness \citep{meteors}. We can expect fluffy grains ejected by the disc wind to shatter during collisions. The resulting products of collisions are then deposited into the colder parts of the disc and integrated into the cometary dust. This  process has been implied by the presence of thermally processed crystalline grains in the solar system comets \citep{comets} and by the thermal history of refractory materials in meteorites \citep{meteorites}, but it has also been observed as a transient outflow of crystalline dust during a stellar outburst \citep{EXLup}.

\section{Conclusions}
\label{SecConclude}

In this work we expanded on our exploration of the influence of radiation pressure on the dust dynamics in protoplanetary discs. Since the strength of this force is dictated primarily by the stellar luminosity, we first looked at an example of Herbig Ae star and noticed several impacts on the dynamical evolution of dust:
\begin{itemize}
    \item[--] Radiative transfer tells us that big grains ($\gtrsim$1$\mu$m) can survive the closest to the star, but strong radiation pressure forces impose further selection effects on the grain sizes. Depending on the stellar luminosity, the dust populating the inner disc edge could be tens of microns or more in size as smaller ones are removed by radiation pressure. 
    \item[--] A strong radiation pressure force easily and quickly ejects any smaller dust that might appear on the disc surface due to the inner turbulence or disc inhomogeneities. This outflow might appear as episodic dust ejections.
    \item[--] Big grains have grey opacity for both the stellar and the near infrared radiation. Grey opacity leads to a non-radial radiation pressure force originating from the hot dust itself. The force is the strongest at the inner edge where dust is the hottest. Its effect on the dust ejection is significant as it lifts dust out of the disc surface and exposes it to the strong stellar pressure that ejects it far away into the colder parts of the disc. 
    \item[--] The non-radial force component continuously pushes the inner edge dust upward, counterbalancing the gas drag force and dust settling. This process is independent of the geometric curvature of the inner rim because its source is the near infrared radiation from the surface itself. Dust trapped in such a force limbo can stay exposed to thermal processing and dust growth longer than usual.
    \item[--] In general, radiation pressure opposes dust accretion within the inner disc surface layers, which leads to dust accumulation at the optically thick inner disc edge.
\end{itemize}

In T Tau stars the luminosities are lower and, therefore, the stellar radiation pressure has less dramatic effects. Nonetheless, it still should not be ignored, as we already demonstrated in our previous work where we combined MHD simulations and dust radiative transfer \citepalias{Vinkovic21}. Now we have two improvements - we incorporate time changes in the MHD density and velocity, and we implement the non-radial radiation pressure force. After exploring examples of dust trajectories driven by gravity, gas drag, disc wind, and radiation pressure around a T Tau star, we reached several conclusions:
\begin{itemize}
    \item[--] The radiation pressure force is still effective in opposing dust accretion at the inner disc edge. The effect is less dramatic than in Herbig Ae stars, since here grains of a few micrometres or less are affected. Such grains tend to stop when accretion brings them to the surface of optically thick inner disc edge. 
    \item[--] Once the accreting grains lose their inward radial velocity, they undergo two types of trajectories. Grains at smaller disc heights settle toward the midplane where a gas backflow can take them further back into the disc. Grains at larger disc heights are influenced by the non-radial radiation pressure force, which pushes them upward and then back into the inner disc surface. It is a form of convective motion, which can expose dust to more complex physical and chemical processing within this hot disc region. 
    \item[--] Further away from the inner dusty disc edge, the stellar magnetic field meets the disc field and a current sheet forms in that contact zone. We see that the dusty disc surface is significantly disrupted at the bottom of the current sheet. This disruption stirs up the dust and enables dust to move above the disc.
    \item[--] Our MHD model is not precise enough to pinpoint the exact position of the current sheet and its disc interaction, but we capture its erratic movement and how this affects the dust. The gas density column is formed with the current sheet, anchored at the disc surface. Without radiation pressure, the dust captured by that column can be lifted up, tossed around by the gas column movements, and then thrown either inward or outward. However, when the inevitable radiation pressure is included, the dust ends up pushed away from the star, further back into outer disc regions. 
    \item[--] Smaller grains are more easily captured by the current sheet gas and can reach significant heights above the disc. We expect this zone to be a source of irregular dust outflows. Radiation pressure plays a significant role in the dynamics of such outflows. 
\end{itemize}

Our main conclusion is that the inner regions of optically thick dusty discs cannot be properly studied without understanding how the radiation pressure force impacts the dust distribution and dynamics. One must keep in mind that our model is a simplified version of reality. Dust dynamics is most probably affected by various other forces in realistic discs. Nonetheless, what remains certain is that radiation pressure, including its non-radial version at the hot inner edge, interferes with dust accretion and settling, and can cause dust outflows in combination with gaseous disc winds.

\section*{Acknowledgements}

Authors acknowledge collaboration with the Croatian project STARDUST through HRZZ grant IP-2014-09-8656. M\v{C} developed the setup for star-disc simulations while in CEA, Saclay, under the ANR Toupies grant. His work in NCAC Warsaw is funded by the Polish NCN grant No. 2019/33/B/ST9/01564, and in Opava by the Czech Science Foundation (GA\v{C}R) grant No.~21-06825X. We thank ASIAA, Taiwan and CAMK, Poland, for access to Linux computer clusters used for the high-performance computations, and the {\sc pluto} team for the possibility to use the code. DV acknowledges Technology Innovation Centre Me\dj{}imurje for logistical support.

\section*{Data Availability}

The dust dynamics calculations in this paper were created using the pseudo-codes described in Sections \ref{Sec:HAe_trajectory} and \ref{Sec:TTau_trajectory}. Our Python implementation of these pseudo-codes will be shared on reasonable request to the corresponding author. The algorithm requires import of simulation results from MHD simulations. For this we used already published results from previous research as described in Section~\ref{SubSec_disccorona}. The non-radial radiation pressure vector field used in our models is available in the supplementary materials (see Appendix \ref{AppendixAnima}).






\section*{Supporting Information}

Supplementary data are available at MNRAS online

\noindent \texttt{\bf Figure\_4.mp4}

\noindent \texttt{\bf Figure\_5\_density.mp4}

\noindent \texttt{\bf Figure\_5\_velocity.mp4}

\noindent \texttt{\bf Figure\_7.mp4}

\noindent \texttt{\bf Figure\_8.mp4}

\noindent \texttt{\bf Figure\_9.mp4}

\noindent \texttt{\bf Figure\_10.mp4}

\noindent \texttt{\bf Figure\_11.mp4}

\noindent \texttt{\bf Figure\_12.mp4}

\noindent \texttt{\bf Figure\_13.mp4}

\noindent \texttt{\bf Radiation\_pressure\_code.pdf}

\noindent \texttt{\bf LELUYA\_Radiation\_Pressure.dat}

\noindent Please note: Oxford University Press is not responsible for the content or functionality of any supporting materials supplied by the authors. Any queries (other than missing material) should be directed to the corresponding author for the article.


\appendix

\section{MHD animations, dust grain dynamics animations, and radiation pressure vector field}
\label{AppendixAnima}

We include as supplementary material a collection of animations showing the MHD wind model and dust flows. We also provide a 2D array data of the non-radial radiation pressure vector, calculated with the code LELUYA (http://www.leluya.org), and a Python code on how to read it.

The animations include the following:
\begin{itemize}
\item[--] \texttt{Figure\_4.mp4} showing dust outflow in a disc around a Herbig Ae star (see Fig.~\ref{fig:HAeFig4}).
\item[--] \texttt{Figure\_5\_density.mp4} and \texttt{Figure\_5\_velocity.mp4} showing gas density and velocity in our MHD model (see Fig.~\ref{fig:MHD}).
\item[--] \texttt{Figure\_7.mp4} showing the flow of dust grains at the inner edge of optically thick dusty disc around a T Tau star when we ignore the radiation pressure force (see Fig.~\ref{fig:Inner4micNOradPress}).
\item[--] \texttt{Figure\_8.mp4} is the same as previous animation, but this time with included stellar radiation pressure force (see Fig.~\ref{fig:Inner4micNonRadialOff}).
\item[--] \texttt{Figure\_9.mp4} is the same as previous animation, but this time with the complete radiation pressure force description, including the non-radial diffuse contribution (see Fig.~\ref{fig:Inner4mic}).
\item[--] \texttt{Figure\_10.mp4} showing interaction between the MHD gas wind and the dusty disc when the radiation pressure force is ignored (see Fig.~\ref{fig:05noPress}).
\item[--] \texttt{Figure\_11.mp4} is the same as previous animation, but this time with included stellar radiation pressure force (see Fig.~\ref{fig:05withPress}).
\item[--] \texttt{Figure\_12.mp4} is the same as previous animation, but with a smaller grain size (see Fig.~\ref{fig:01withPress}).
\item[--] \texttt{Figure\_13.mp4} is the same as previous animation, but without the radiation pressure force (see Fig.~\ref{fig:01NoPress}).
\item[--] \texttt{Radiation\_pressure\_code.pdf} is an example of a Python code to read and plot the non-radial radiation pressure force (see Fig.~\ref{fig:HAeFig1} and Fig.~\ref{fig:Inner4mic}).
\item[--] \texttt{LELUYA\_Radiation\_Pressure.dat} is a 2D array of the non-radial radiation pressure force vector stored in spherical coordinates.

\end{itemize}

\bsp	
\label{lastpage}
\end{document}